\documentclass[reprint,
superscriptaddress,
frontmatterverbose, 
amsmath,amssymb,
aps,
prb,
floatfix,
showkeys
]{revtex4-2}

\usepackage{float}
\usepackage{graphicx}
\graphicspath{{./}}
\usepackage{dcolumn}
\usepackage{bm}
\usepackage[left=2cm,right=2cm,top=2cm,bottom=2cm]{geometry}
\usepackage{mathtools}
\usepackage{amsmath}
\usepackage{xcolor}
\usepackage{soul}
\usepackage{multirow}
\mathchardef\hyphen="2D

\begin{document}

\preprint{APS/123-QED}

\title{Temperature Dependence of Phonon Energies and Lifetimes in Single- and Few-layered Graphene}

\author{Markos Poulos}
 \email[Corresponding Author: ]{markos.poulos@insa-lyon.fr}
 \affiliation{INSA Lyon, CNRS, CETHIL, UMR5008, 69621 Villeurbanne, France}

\author{Konstantinos Papagelis}
\affiliation{Aristotle University of Thessaloniki, School of Physics, 54124 Thessaloniki, Greece }
\affiliation{Institute of Chemical Engineering Sciences, FORTH/ICE-HT, 26504 Patras, Greece}
 
\author{Emmanuel N. Koukaras}
\affiliation{Institute of Chemical Engineering Sciences, FORTH/ICE-HT, 26504 Patras, Greece}
\affiliation{Aristotle University of Thessaloniki, School of Chemistry, 54124 Thessaloniki, Greece }

\author{George Kalosakas}
\affiliation{University of Patras, Department of Materials Science, GR-26504 Rio, Greece}

\author{Giorgia Fugallo}
\affiliation{CNRS, Université de Nantes, LTEN, UMR6607, 44306 Nantes, France}

\author{Konstantinos Termentzidis}
\affiliation{CNRS, INSA Lyon, CETHIL, UMR5008, 69621 Villeurbanne, France}

\begin{abstract}
In this work, we have studied the phonon properties of
multi-layered graphene with the use of Molecular Dynamics (MD) simulations and the
k-space Autocorrelation Sequence (k-VACS) method. We
calculate the phonon dispersion curves, densities of states and lifetimes $\tau$ of few-layered graphene of 1-5 layers and graphite. $\Gamma$-point phonon energies and lifetimes are investigated for different temperatures ranging from 80 K to 1000 K. The study focuses on the impact of the interlayer interaction and temperature on the energies and lifetimes of the $\Gamma$-point phonons, as well as the type of interlayer potential used. For the later we used the Kolmogorov-Crespi (KC) and the Lennard-Jones (LJ) potentials. We have found that the number of layers $N$ has little effect on the intra-layer (ZO and G) mode energies and greater effect on the inter-layer (Layer Shearing and Layer Breathing) modes, while $\tau$ is generally affected by $N$ for all modes, except for the Layer Shear mode. The trend of $N$ on the lifetimes was also found to independent of the type of potential used. For the Raman-active G phonon, our calculations show that the lifetime increase with $N$ and that this increase is directly connected to the strength of the interlayer coupling and how this is modelled.
\end{abstract}

\maketitle

\section{\label{sec:Intro}Introduction:}

Graphene since its discovery has attracted widespread attention due to its exceptional properties~\cite{Urade2023}, especially for its excellent electronic conductivity~\cite{Novoselov2004}, its performance in optoelectronic devices~\cite{Wang2019}, its mechanical strength~\cite{Lee2018} and its high thermal conductivity~\cite{Sang2019}. This increased interest arised from the fact that high-quality monolayer graphene (1L) can be produced by means of mechanical exfoliation of Bulk Graphite (BG), as per the famous scotch-tape experiment~\cite{Novoselov2005}. There has also been recent interest in few-layered graphene (FLG), as it also exhibits interesting properties, ranging from high electrical~\cite{Murata2019} and thermal conductivity~\cite{Singh2011, Liu2016, Hamze2020}, to enhanced lubricity~\cite{Fu2019,Androulidakis2020}  which is makes it an ideal filler material in composites for thermal management applications~\cite{Shahil2021}, while most recently, unconventional superconductivity was discovered in twisted bilayer graphene~\cite{Cao2018}. All the above properties are strongly phonon-related, as electrical conductivity is hypothesized to be phonon-limited~\cite{Davis2023}, while the thermal conductivity of graphene is almost exclusively due to phonons~\cite{Nika2011,Ghosh2008}. It is therefore of interest to map the change of the phonon properties of graphene as a function of the number of layers $N$, as well as of temperature.

In monolayer graphene (1LG), besides the 3 acoustic modes (longitudinal LA, transverse TA and out-of-plane ZA) there exist 2 optical modes, the out-of-plane ZO and the in-plane LO and TO modes which at the $\Gamma$-point become the doubly-degenerate E$_{2g}$ mode, also known as the G-mode due to the Raman G-band that it is responsible for. The presence of multiple layers modifies the phonon structure of graphene, as new modes appear: existing intralayer optical modes split into Davydov multiplets, and new interlayer (IL) modes appear due to the folding of the Brillouin zone edge acoustic phonons~\cite{Lindsay2011, Nemanich1977}. These emergent interlayer modes are categorised as layer Shear and layer Breathing modes, where the layers move as units in the in-plane and the out-of plane direction, respectively, with a phase difference between each other.  The Layer Breathing mode (LBM) can also be found noted as ZO$'$, while the Shear mode is also denoted as C-mode, again due to the Raman C-band signal.

Davydov splitting, also known as factor group splitting, refers to the removal of degeneracy of internal degrees of freedom due to weak external interactions, in materials whose unit cell is composed of replicated units, such as molecular crystals and layered Van der Waals (VdW) materials~\cite{Davydov_Excitons, Ghosh1976}. The resulting modes usually have weakly split energies ($\Delta\omega<$ 5~cm$^{-1}$) and not far from the unperturbed system~\cite{Molina2015}, although in cases of strong interlayer interaction such as in layered 2D antimonene, large splits ($\sim$70~cm$^{-1}$) have been reported~\cite{Gibaja2016}. As a result, in N-layered graphene there exist (N-1) Shear, (N-1) Breathing, N ZO and N G phonons. For a more detailed analysis of the phonon modes of few-layered graphene, we refer the interested readers to the work of Saha et al.~\cite{Saha2008}.

The dependence on temperature $T$ of the phonon energies $\omega(T)$ and lifetimes $\tau(T)$ mainly arises due to anharmonicity in the interatomic interactions. As has been long established by Maraduddin and Fein (1962), by evaluating the phonon self-energy $\Sigma(\omega)$ of an interacting phonon   
system~\cite{Maradudin1962}, one can obtain from its real $\Delta(T)$ and imaginary part $\Gamma(T)$, respectively, the anharmonic phonon frequency shift and finite linewidth in the energies $\omega_0$ of the harmonic non-interacting phonons. The T-dependence then mainly arises due to the dependence of $\Gamma$ and $\Delta$ on the phonon populations $n(\omega, T)$. 
Lattice thermal expansion also contributes to $\Sigma(\omega)$ by the implicit inclusion of higher order anharmonic terms in the renormalization of the harmonic energy~\cite{Hellman2013} leading to a temperature-dependent $\omega_0 (T)$ since at higher temperatures these terms can no more be considered as small perturbations~\cite{Gu2019}. However, if one can determine the lattice thermal expansion independently (e.g. experimentally), the
shift at a given temperature can be obtained computing the harmonic frequency at the lattice parameters corresponding to that temperature.

The phonon energies of layered graphene and graphite have been experimentally probed quite extensively in the literature, mostly concerning the zone-center Raman-active G phonon ($\omega_G\approx1580$~cm$^{-1}$).  For graphite, the phonon dispersion curves by means of Inelastic X-ray Scattering (IXS) have been reported~\cite{Maultzsch2004}, and for 1L graphene studies on the T-dependence of $\omega_G (T)$ by means of conventional Raman spectroscopy are also present in the literature, with excellent agreement between them~\cite{Sullivan2017,Calizo2007, Han2022}. Studies of the effect of the number of layers $N$ on the phonon energies of FLG have also been conducted~\cite{Ferrari2006,Late2011,Tan2012, Lui2012, Lui2014}

For the experimental study of the phonon lifetimes of 1L graphene, conventional Raman spectroscopy is usually performed on suspended samples in order to exclude the effect of the substrate~\cite{Lin2011,Han2022}. The lifetime of a Raman-active mode can be estimated by the linewidth $\gamma$ in cm$^{-1}$ as $\tau= \left(2 \gamma \pi c \right)^{-1}$~\cite{Katsiaounis2021,Kang2010}. The major contributions to the linewidth are due to el-ph and ph-ph interactions $\gamma_G = \gamma_G^{ph-ph} + \gamma_G^{el-ph}$. Most notably for the G-mode, the major contribution to $\gamma_G$ is due to the electron-phonon part~\cite{Chatzakis2011, Wang2010, Han2022}. For the G-mode, the ph-ph contribution can be also probed separately by means of Time-Resolved Incoherent Anti-Stokes Raman Spectroscopy (TRIARS). It appears commonly accepted that for graphite $\tau^{ph-ph}_{G}$ ranges between 2-2.5~ps at room temperature~\cite{Chatzakis2011,Yan2009}. For 1L, however, reported values are dispersed, from $\sim$1-1.5~ps~\cite{Kang2010, Katsiaounis2021} to 2.5~ps~\cite{Wang2010} for supported samples, and as high as 4.9~ps for samples supported on colloidal solutions~\cite{Kumar2009}. The growth method and the sample quality have a big impact in the measured value. An overview of the different measured values reported in the literature can be found in Katsiaounis et al~\cite{Katsiaounis2021}.

Phonon energies and lifetimes are usually studied theoretically by means of \textit{ab initio} methods based on Perturbation Theory. Initially calculations in the literature included only the electron-phonon (e$\hyphen$ph) andthree-phonon (3ph) contributions to the temperature-dependence of $\tau_G(T)$ in single-layer graphene~\cite{Bonini2007}. These calculations showed that the e$\hyphen$ph contribution overwhelmingly dominates the temperature dependence,causing $\gamma_G$ to decrease with temperature. This finding starkly contrasts with all the expermental works cited above. The discrepancy was only recently resolved. It has been reported that the 4ph contribution is actually much more significant than 3ph, although the e$\hyphen$ph contribution still remains dominant~\cite{Han2022}. The same authors reported that phonon renormalization due to lattice thermal relaxation was crucial to avoid overestimating the 4$\hyphen$ph scattering rates~\cite{Han2022}.

An alternative way to calculate phonon properties without the above limitations is to employ the k-space Velocity Autocorrelation Sequence method within classical Molecular Dynamics (MD) simulations, which we have previously employed to obtain the temperature dependence of the $\Gamma$-point phonon energies in SLG~\cite{Koukaras2015}. With this method we can obtain phonon energies and lifetimes for all $k$-points in the Brillouin Zone allowed by the Periodic Boundary Conditions~\cite{Noid1977,Koukaras2015}. For each $k$-point the Spectral Energy Density is obtained, which is then fitted by Lorentzian functions. The positions and widths of the peaks correspond to the phonon energies and linewidths respectively. The Phonon Density of State (PDOS) can be also extracted within the same scheme and from the same data sets. The two main advantages of this method is that it inherently takes into account by default all orders of anharmonicity and lattice expansion, while it is also applicable to both periodic and non-periodic systems. The method has been described extensively in a previous works and readers interested in the details are refered to it~\cite{Koukaras2015}. The further details and a theoretical proof of the method are also provided in Appendices~\ref{app:kVACS_Method} and~\ref{app:kVACS_proof}. 

kVACS has been extensively used in the past to study vibrational properties in condensed matter applications, from liquids~\cite{Rahman1964} to molecules~\cite{Kohanoff1994} and the temperature dependence of phonon modes in Si~\cite{Wang1989}, to surface modes in Cu~\cite{Papanicolaou1995,Dietlevsen1991}. In more recent studies, it has been applied to obtain the phonon spectra of carbon nanotubes~\cite{Thomas2010}, monolayer graphene~\cite{Mafakheri2021} and nanoscale Si~\cite{Chen2019}. Recently it was also used to investigate the contribution of the phonon vibrational structure in the mechanism of thermal rectification of certain nanostructures, as in the example of asymmetric two-phase Si-nanowires~\cite{Desmarchelier2021}, and in MoSe$_2$-WSe$_2$ \cite{Zhang2022} and graphene-hBN Van der Waals lateral heterojunctions~\cite{Chen2020}.

In layered materials, adjacent layers interact with each other through weak Van der Waals (VdW) forces arising from an attractive spontaneous dipole-dipole dispersion interaction~\cite{Born1956} and a repulsive interaction arising from the Pauli repulsion of orbital overlap~\cite{Adams2001}. The attractive part is long-ranged, while the repulsive part is short-ranged, and each term is dominant in their respective distance range~\cite{Adams2001}. In MD simulations this interaction has been modelled quite extensively with the classical Lennard-Jones (LJ) potential, however, due to its being a two-body radial potential, it has been found to be too smooth to describe corrugation and variations in the stacking alignment of layered graphitic materials~\cite{Kolmogorov2005}. For this reason, a new type of interlayer potential for graphitic materials was proposed by Kolmogorov and Crespi (herby denoted as KC), which is a registry-dependent potential that takes specifically into account the directionality of the repulsive overlap interaction between $\pi$-orbitals of adjacent layers~\cite{Kolmogorov2005}. This is the main interlayer potential employed in this work. Other registry-dependent potentials have also been developed ever since, such as the ILP potential which includes Coulomb interactions for polar layered materials like hBN~\cite{Leven2014}, the DRIP potential which includes dihedral interactions for twisted bilayer graphene~\cite{Wen2018} and the Lebedeva potential which includes a term dependent on the vertical distance of the $p_z$ orbitals of adjacent atoms, to better describe the compressibility of graphite~\cite{Lebedeva2012}.

The rest of the article is organised as follows: First, a synopsis of the computational methods and interatomic potenials that we used in this paper is made, followed by the discussion of the results. There, we present the calculated phonon dispersion curves and Density of States for Bulk Graphite, and the temperature dependence of the energies and lifetimes of $\Gamma$-point phonons 1-2-3-5 layered graphene and graphite, calculated with the KC potential and compared with all available experimental data. The calculations of energies and lifetimes are repeated using the LJ potential and the trends are compared to the results with KC. Finally the effect of the interlayer coupling strength of the LJ potential on the lifetime of the G phonon is investigated for 2-layered graphene and graphite.\@

\begin{figure*}[h!tbp]
    \begin{center}
        \includegraphics[width=0.9\textwidth]{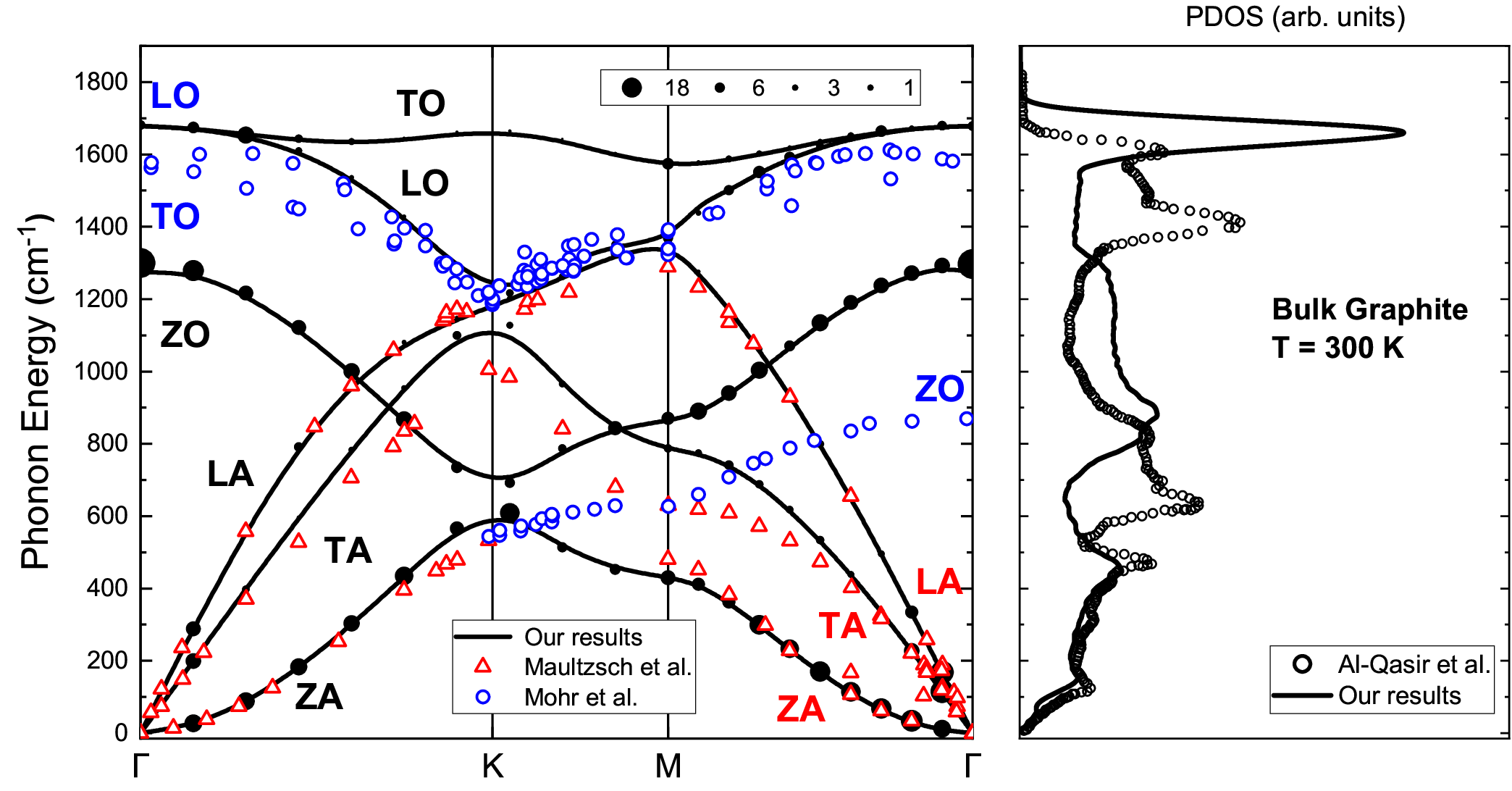}
        \caption{\label{fig:Disp_PDOS} (Left) The calculated phonon dispersion curves of Bulk Graphite (full symbols) at $T=300 \ K$, together with data from IXS measurements (open symbols)~\cite{Maultzsch2004, Mohr2007}. The magnitude of the kVACS symbols is a measure of the calculated lifetime. Straight lines connecting points are b-splines serving as visual guides. (Right) The calculated phonon DOS for Bulk Graphite at $T=300 \ K$ (straight lines), along with data from INS measurements~\cite{AlQasir2020} (open symbols). All PDOS are normalized to unit area. }
    \end{center}
\end{figure*}

\section{Methodology: \label{sec:Method}}
In this work we have performed MD simulations on single- and few-layered pristine graphene (FLG) and graphite (BG), using the kVACS method. The phonon properties of the above systems were studied under temperatures ranging from 80 K to 1000 K in order to obtain the temperature dependence of the phonon energies and lifetimes. 

The interatomic C-C interactions within the same layer were modelled with the reparametrised Tersoff potential by Broido \textit{et al}~\cite{Broido2010}. It shall be pointed out that in a previous work the LCBOP and the AIREBO potentials were shown to give a very poor temperature dependence of the energy of the G phonon in 1L graphene~\cite{Koukaras2015}. A Machine-Learning based Gaussian Approximation Potential for graphene has also been reported to give very promising results and reproduced very well the DFT dispersion curves and thermal expansion coefficients~\cite{Rowe2018}. However the increased computational cost of such potentials was prohibitive for the purposes of our work, which mainly focus on probing the interlayer interactions, and which are independent on the intralayer potential used. For the above reasons, the revised Tersoff potential was deemed the best solution. For the interlayer interactions we used the reparametrised full Kolmogorov-Crespi (KC) potential with a taper function as described in~\cite{Ouyang2018} and whose form is given in Appendix~\ref{app:KC}. In order to investigate the impact of the nature of the interlayer potential and the effect of the coupling strength, we repeated part of the calculations using the classical Lennard-Jones (LJ) potential for comparison, given by
\begin{eqnarray}
    V_{LJ}(r) = 4\epsilon \left[\left(\frac{\sigma}{r}\right)^{12} - \left(\frac{\sigma}{r}\right)^6\right]
\end{eqnarray}
For the KC potential, a global cutoff of r$_c$ = 12 $\AA$ was chosen. For the LJ potential, the parameters $\epsilon$= 4.6 meV and $\sigma$ = 3.276 $\AA$ used were taken from Lindsay \textit{et al}~\cite{Lindsay2011}. In all of our calculations, all layers can interact with all the others, which mostly happens via the long-range part of the VdW interaction, which "passes through" layers. As it has been underlined in previous works, in few-layered graphene the cleavage energy (CE) (the energy required to split a bulk material in half along a basal plane) and the interlayer Binding Energy (BE) (the energy required to separate a bulk material into intdividual non-interacting layers) is different, which is a sign of interlayer interactions longer than between adjacent layers~\cite{Gould2013}. Experimental measurements of CE and the BE of graphite support this claim, as their difference was found to be $\sim$15\%~\cite{Wang2015}.

The FLG systems studied were 1L, 2L, 3L, 5L graphene and BG, all in Bernal (-AB-) stacking. The computational cell used consisted of 40x40 primitive cells with Periodic Boundary Conditions (PBC), having sides of 1 nm. Each system was first structurally and thermally relaxed for every temperature step: an initial conjugate gradient (CG) static energy minimization followed by a 125 ps NVE initial equilibration, a 250 ps NPT structural relaxation and a final 125 ps NVT thermal equilibration. 

During the production step to obtain the kVACS spectra, 10 independent equilibrium NVE runs with randomized initial velocities were performed. This statistical sampling was done in order to obtain good ensemble averages and to ensure ergodicity. For the simulations using KC as the interlayer potential, each run consisted of 655.360 timesteps or 327.68 ps, for a total of 3.28 ns of production time per studied case. The sampling rate of the velocity data was every 10 timesteps or 5 fs, giving an $\omega$-resolution of 1.1 cm$^{-1}$. Comparative convergence tests with higher resolution were performed for the 2L case and are presented in Section 2 of the Supplemental Material~\cite{supplementary}. No significant deviation of the calculated $\omega$ and $\tau$ with resolution were found. For the case of the LJ potential, convergence tests showed that this resolution gave poor results. For this reason, for all the LJ results presented in this paper the per-run simulation time was 655.36 ps, double that of KC, and the sampling rate was every 15 steps, giving an $\omega$-resolution of 0.5 cm$^{-1}$. The number and frequency of sampled data points was eventually the result of a compromise between computational efficiency and sufficient $\omega$-resolution. For all simulations, the timestep used was 0.5 fs.  All MD simulations were performed using the LAMMPS code~\cite{LAMMPS}.

The phonon properties were calculated from the kVACS method and the spectra obtained were fitted using simple Lorentzian functions. The Brillouin Zone was sampled along the path $\Gamma$-K-M-$\Gamma$. The Lattice Dynamics (LD) simulations were also performed using LAMMPS's \textit{dynamical\_matrix} command to calculate and output the force constant matrix of the systems under study. The dynamical matrices for specific k-points were calculated by performing a spatial Fourier transform over the equilibrium positions and finally the phonon energies were obtained by the diagonalization of the dynamical matrix.

Finally, an important point should be raised, concerning the resolution of the k-points examined. As we are using finite computational cells and PBC, not any arbitrary k- can be sampled, but only those that are linear combinations of $\frac{1}{N_1}\textbf{b}_1$ and $\frac{1}{N_2}\textbf{b}_2$, where $\textbf{b}_1$ and $\textbf{b}_2$ are the primitive inverse lattice vectors of the respective unit cell and $N_1$, $N_2$ the number of unit cells along each lattice vector, respectively. Here, $N_1=N_2=40$. The procedure to obtain the correct k-point resolution has been extensively documented in a previous study~\cite{Koukaras2015} and for further details we refer the reader to it.

\section{\label{sec:Results}Results and Discussion:}

For all multi-layered systems, as has been mentioned in the Introduction, all optical modes (Shear, LBM, ZO and G) appear in multiplets, therefore a choice of representation must be made. In the rest of this work, the highest-energy ZO Davydov multiplet is chosen to represent the said mode, which corresponds to a vibration where the crystallographically equivalent C atoms of each layer vibrate in phase. For the G mode, the lowest-energy multiplet is chosen, which also corresponds to the in-phase vibration of the equivalent C atoms. This guarantees that the G-mode chosen will be Raman active, independent of the number of layers~\cite{Saha2008}. Concerning the interlayer modes, both for the LBM and the C-mode branches for each $N$ the highest-lying branch corresponds to an anti-phase layer vibration, while the lowest-energy branches correspond to phonons where mostly the outer layers vibrate~\cite{Lui2014}. For reasons of consistency with the experimental studies, we will represent the shear modes by the highest-energy branch (anti-phase), while for the breathing mode we choose the lowest-lying one, where layers oscillate like an accordion. The reason for this choice is that these modes represent an actual layer shearing/breathing vibrational pattern, while also being always Raman-active and giving the highest intensity~\cite{Tan2012}. An important remark should be made about the LBM of graphite, which doesn't follow the aforementioned vibrational pattern, as all layers vibrate out-of-phase. A visual illustration of the vibrational patterns of all the modes mentioned above is provided in Figs. SM-1 and SM-2 of the Supplemental Material~\cite{supplementary}.

From a first comparison of the results, it turns out that the number of layers $N$ affects differently the intralayer (acoustic, ZO and G) and the interlayer (C and LBM) modes accross the entire Brillouin zone and throughout the entire temperature range. Specifically for the intralayer modes, the calculated energies increase slightly going from from 1L to FLG and BG in the order of 1-5 cm$^{-1}$, therefore no visible difference was seen between the equivalent phonon dispersion curves for different $N$ and different T, especially at the energy scale covering all the mode energies (0-1700 cm$^{-1}$). For this reason, the dispersion curves of graphite were thus chosen to represent the intralayer curves of FLG for all $N$, which we present in Fig.~\ref{fig:Disp_PDOS}, along with information about the calculated lifetimes, in comparison with data from Inelastic X-Ray Scattering (IXS) measurements~\cite{Maultzsch2004,Mohr2007}. We also provide the calculated Density of States (PDOS) of graphite, compared with IXS measurements from a different author~\cite{AlQasir2020}, all at T=300~K. The interlayer modes are affected in a more intricate way and further details will be provided later in the text. A comparison of all phonon branches for all $N$ is also provided in Fig. SM-8 of the Supplemental Material~\cite{supplementary}.

We directly observe from Fig.~\ref{fig:Disp_PDOS} that the acoustic modes are described with very good accuracy by the Tersoff-2010 potential with respect to experimental data, as was the case in the initial work of potential reparametrization by Broido \textit{et al}~\cite{Broido2010}, where the dispersion curves were obtained by simple LD. Concerning the calculated lifetimes, they fall within the range of $\sim$1-20~ps and two principal remarks are to be made. First of all, the out-of-plane intralayer modes (ZA and ZO) have noticeably much larger lifetimes than the in-plane ones (TA, LA, LO, TO), ranging between $\sim$5-20~ps and increasing close to the $\Gamma$-point. Interestingly, the ZA and ZO mode lifetimes have close values. This is in stark contrast with the in-plane modes, where the acoustic ones globally have larger lifetimes (5-10~ps) than optical ones (1-5~ps).

The energies of the LO/TO branches are overstimated by $\sim$10\%, while the ZO branch is significantly overestimated at about 45\%. Also, the description of the LO and TO branches is satisfactory on the $\Gamma$-M path (armchair), but significantly overestimates about 27\% the splitting of the LO branch at the $\Gamma$K (zig-zag) and KM (parallel to the C-C bond axis) paths. Also, the TO branch has higher energies than the LO branch, which has also been observed in~\cite{Koukaras2015} and can be attributed to the Tersoff potential. It should be further noted that the discrepancy in the energies the optical modes shall not be attributed to the kVACS method, since even in the initial work, the dispersion curves for the optical modes, obtained by Lattice Dynamics, exhibited the same level of discrepancy with experiment~\cite{Broido2010}. Besides, as it will be later shown, the relative difference between LD and kVACS energies of the ZO and G phonons is of the order of 1\%. The original authors attributed the failure of the simultaneous fitting of both optical and acoustic modes to the short range of the potential~\cite{Broido2010}.

The PDOS calculated with kVACS is compared with Inelastic Neutron Scattering (INS) measurements for graphite~\cite{AlQasir2020}, both presented in Fig.~\ref{fig:Disp_PDOS} (right). Both calculated and experimental PDOS are normalized to unit area. We attribute the differences with experimental data to the optical modes, since from the comparison of the Dispersion Curves in Fig.~\ref{fig:Disp_PDOS} (left), the acoustic modes are perfectly matching the experimental data. Also, in our calculations no account of the out-of-plane direction ($\Gamma$-A) of the Brillouin Zone of graphite was taken, those modes are therefore missing from the PDOS.

\begin{figure}[!htb]
    \includegraphics[width=0.85\columnwidth]{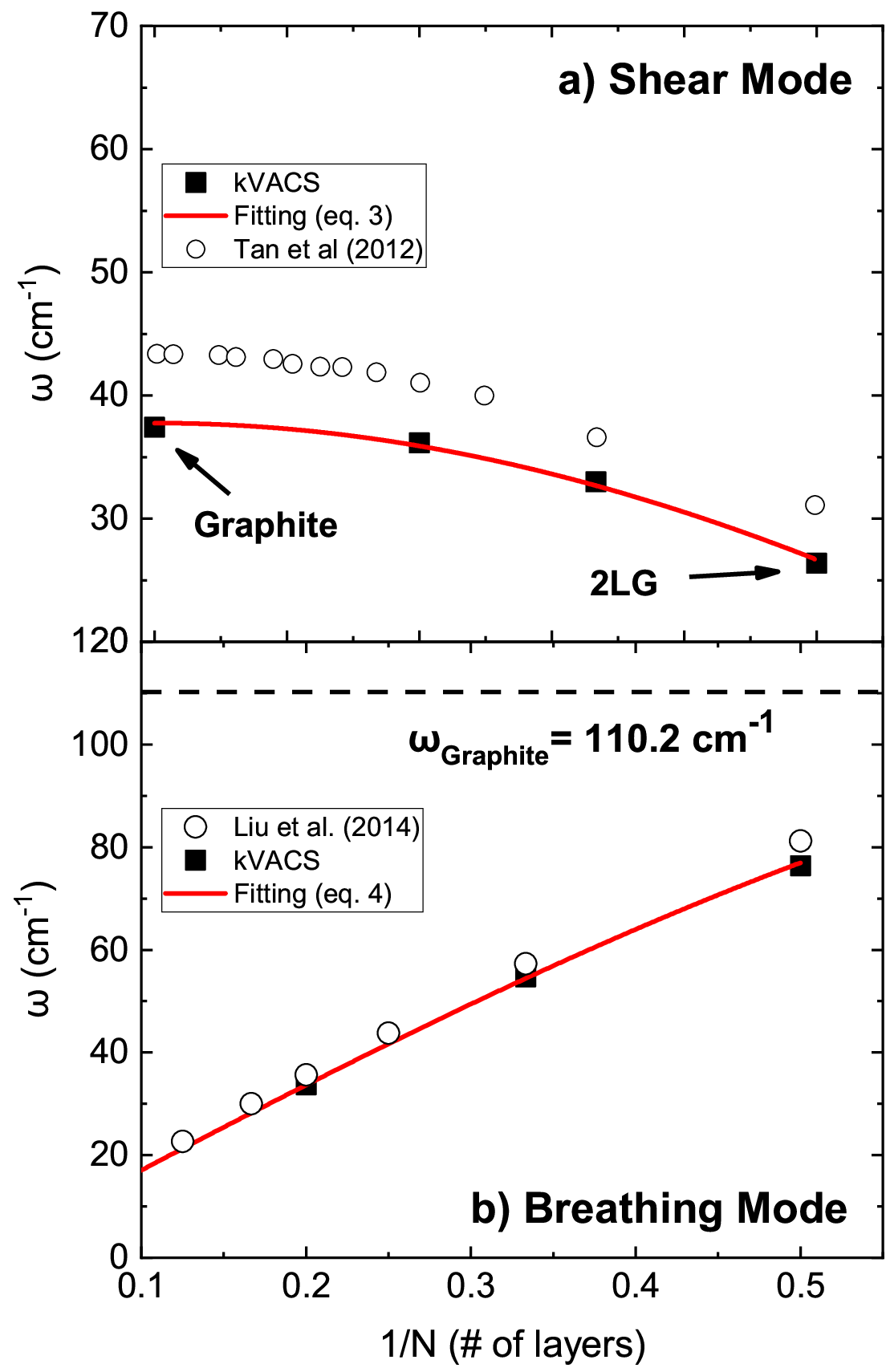}
    \caption{\label{fig:Layers_Dep} The calculated dependence of the FLG (a) Shear and (b) Breathing mode at $T=300 \ K$ on the number of layers $N$ (full symbols), along with the fitting of the energies as given by eqs. (\ref{eq:Shear_Linear_Chain}) and (\ref{eq:Breathing_Linear_Chain}) (straight line). Experimental measurements by Tan et al. \cite{Tan2012} and Lui et al \cite{Lui2014} respectively are also shown for comparison (hollow symbols).}
\end{figure}

It is also of interest to investigate the effect of $N$ on the energy of the interlayer phonon modes. For this, in Figs~\ref{fig:Layers_Dep}a) and~\ref{fig:Layers_Dep}b) we plot the energy of the the true Shear and Breathing modes at T=300 K, as a function of the $N$. As it can be clearly seen, the two modes have opposite trends, as the C-mode hardens with increasing $N$, while the LBM softens, as is also attested by the experimental measurements. This trend can be explained by a simple linear chain model, where layers are modelled by single particles and the interlayer mode frequencies are given by~\cite{Pizzi2021}

\begin{eqnarray}
    \omega^{(i,N)}_{\nu} &=& \sqrt{ \frac{2k_{\nu}}{\mu} \left\{1-\cos\left( \frac{\left(i-1\right)\pi}{N}\right) \right\} }  \label{eq:Linear_Chain}
 \end{eqnarray}
\noindent where $N$ is the number of layers present, $\nu=1,2,3$ represents the interlayer mode branch ($\nu=1,2$ the doubly-degenerate C-mode and $\nu=3$ the LBM), of $N$ modes each, indexed by $i=1,2,3 \dotsc N$ ($i=1$ refers to rigid translation). Also, $k$ is a measure of the force constant per unit area that corresponds to each vibration (shear force for the C-mode and tensile force along the c-axis for the LBM) and $\mu$ is the mass per unit area of a single layer, taken equal to $\mu = 7.6 \times 10^{-27}$ kg$\cdot\AA^{-2}$ . The model has been succesfull to capture the trends in experimental data for the Shear \cite{Lui2014} and Breathing modes \cite{Tan2012}, by simply setting $i=N$ and $i=2$ in eq. (\ref{eq:Linear_Chain}) respectively. After some manipulation, these energies, measured in cm$^{-1}$ become

\begin{eqnarray}
    \omega_{C}(N) &=& \frac{1}{\pi c} \sqrt{\frac{\alpha}{\mu}} \cos\left( \frac{\pi}{2N}\right) \label{eq:Shear_Linear_Chain}
 \end{eqnarray}

 \begin{eqnarray}
    \omega_{LBM}(N) &=&  \frac{1}{\pi c} \sqrt{\frac{b}{\mu} } \sin \left( \frac{\pi}{2N}\right)
    \label{eq:Breathing_Linear_Chain}
 \end{eqnarray}

\noindent In eqs. (\ref{eq:Shear_Linear_Chain}) and (\ref{eq:Breathing_Linear_Chain}) , $\alpha$ and $b$ are different symbols used to represent $k_{\nu}$ for the shear and tensile force constants respectively, and $c=2.997 \times 10^{10}$ cm$\cdot$s$^{-1}$ is the speed of light. It should also be noted that in eqs.~(\ref{eq:Shear_Linear_Chain}) and (\ref{eq:Breathing_Linear_Chain}) the pre-factors of the sinusoidals stand for the C-mode and LBM energies of graphite, respectively. Again for the LBM of graphite, this energy does not derive from $ \lim_{N \to \infty} \omega_{LBM}(N)$ of eq.~(\ref{eq:Breathing_Linear_Chain}), since the former refers to an anti-phase vibration and not the ``accordion-like'' motion of the true LBM (see Fig. SM-1 of the Supplemental Material~\cite{supplementary}), which for $N \to \infty$ tends to zero.

\begin{figure*}[!htb]
    \includegraphics[width=0.85\textwidth]{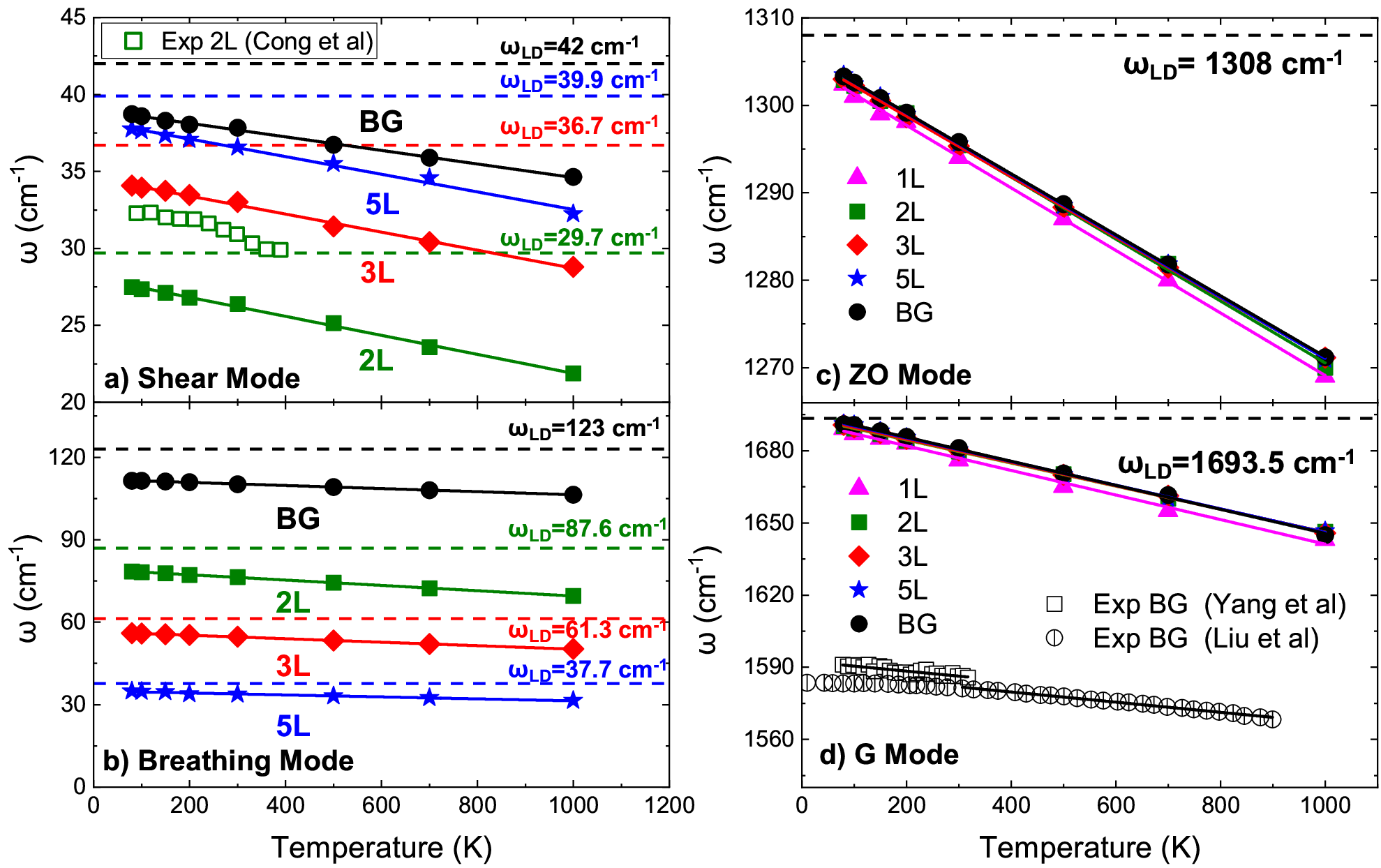}
    \caption{\label{fig:Energy_TDep} The calculated temperature dependence of the true FLG (a) Shear (b) Breathing (c) ZO and (d) G phonon mode energies with the number of layers $N$. The correspondence of symbols with $N$ is as follows. 2L: green squares, 3L: red diamonds, 5L: blue stars, BG: black circles. The hollow symbols represent experimental data by Cong et al~\cite{Cong2014} Yang \textit{et al}~\cite{Yang2020} and Liu \textit{et al}~\cite{Liu2019}. The straight lines with the same color correspond to the least-square fittings of the associated data. The dashed lines represent the LD energies for each $N$.}
\end{figure*}

From the comparison with the model and the experimental data, we can see that the kVACS simulations follow the model in an excellent way. Namely, for graphite $\omega_{C}^{calc}=37.8$ cm$^{-1}$ and $\omega_{C}^{exp}=43.4$ cm$^{-1}$, while the fitting of the C-mode energies using eq.~(\ref{eq:Shear_Linear_Chain}) gives a shear force of $\alpha_{calc}=0.96 \times 10^{19}$ N$\cdot$m$^{-2}$ for the interlayer interaction, which is very close to the value of $\alpha_{exp}=1.27 \times 10^{19}$  N$\cdot$m$^{-2}$ obtained from the experimental fitting~\cite{Tan2012}. Accordingly, the trend of the LBM energies follows very closely the model of eq.~(\ref{eq:Breathing_Linear_Chain}), as it predicts an energy of $\omega_{LBM}^{calc}=108.8$ cm$^{-1}$ for graphite which is very close to the experimental value $\omega_{LBM}^{exp}=114.6$ cm$^{-1}$. Again by using eq.~(\ref{eq:Breathing_Linear_Chain}), we obtain $b_{calc}=7.97 \times 10^{19}$ N$\cdot$m$^{-2}$, while $b_{exp}=8.84 \times 10^{19}$ N$\cdot$m$^{-2}$. Interestingly, if we calculate the same interlayer phonon energies for graphite using Lattice Dynamics with the same potential parameters, which completely ignores thermal motion and anharmonicity, we obtain $\omega_{C}^{LD}=42$ cm$^{-1}$ and $\omega_{LBM}^{LD}=123$ cm$^{-1}$, which give $\alpha_{LD}=1.19 \times 10^{19}$ N$\cdot$m$^{-2}$ and $b_{LD}=10.18 \times 10^{19}$ N$\cdot$m$^{-2}$.

From the above discussion we can extract two useful conclusions. First, even though the KC potential is a complex multi-parameter potential, and despite the inclusion of thermal motion and anharmonicity within kVACS, the behaviour of the interlayer phonon energies eventually follows a simple linear chain model, with a single adjustable parameter, the interlayer force constants $k_{\nu}$. Secondly, because of the latter argument, a significant part of the discrepancy with experiment can be tracked down to the parametrization of the KC potential. This is also supported by the LD calculations, which, especially for the case of the LBM, reveal that thermal motion and anharmonicity can significantly reduce the effective interlayer force constants and phonon energies. This is more or less expected, as the parametrization of the KC potential used in this work was done via DFT, a zero-temperature method~\cite{Ouyang2018}. 

In the following, we obtained the temperature dependence of phonon energies and lifetimes for the Shear, Breathing, ZO and G modes of FLG. For temperatures higher than the mode-specific Debye temperature $\Theta_D = \hbar \omega /k_B$, where $\omega$ is the harmonic energy of the phonon in question, $\omega(T)$ is linear or quadratic (classical limit), depending of whether the dominant anharmonic process is the 3- or 4-phonon~\cite{Balkanski1983}. As MD obeys classical statistics, though, this behaviour is expected to be present even at low temperatures, due to the Equipartition of energy being valid at all temperatures in Boltzmann statistics, but only at temperatures $T \gg \Theta_D$ for Bose statistics.  Therefore the slopes $\chi_T=\frac{d\omega}{dT}$ are a characteristic measure of phonon anharmonicity. Another possibly interesting metric that can quantify the effect of the thermal motion on the phonon softening could be the relative difference between the LD energy value and the lowest-temperature kVACS value, that is $\delta\omega_{LD}=(\omega_{LD}-\omega_{T=80 K})/\omega_{T=80 K}$. The slopes $\chi_T$ for all $\Gamma$-point phonon modes and numbers of layers are presented in Table~\ref{tab:energy_slopes}.

\begin{table}[h]
    \caption{\label{tab:energy_slopes} The energy slopes with temperature $\chi_T$ for each $\Gamma$-point phonon mode and for various numbers of layers calculated via kVACS. Experimental slopes are taken from~\cite{Yang2020,Liu2019}. All values are given in cm$^{-1}$K$^{-1}$}
    \begin{ruledtabular}
        \begin{tabular}{ccccc}
        \multirow{2}{*}{\# of Layers} &\textbf{Shear}  & \textbf{LBM} \ & \textbf{ZO} \ & \textbf{G} \ \\
        & \multicolumn{4}{c}{\textit{($\times$10$^{-3}$ cm$^{-1}$K$^{-1}$)}}\\\hline
         1& -- & -- & -36 & -51 \\
         2& -6.2 & -9.7 & -35 & -47 \\
         3& -5.9 & -6.3 & -34 & -48 \\
         5& -5.7 & -3.6 & -35 & -49 \\
         BG& -4.4 & -5.6 & -35 & -50 \\
         Exper.& -- & -- & -- & -21 \\  
        \end{tabular}
     \end{ruledtabular}
\end{table}

In Figs.~\ref{fig:Energy_TDep}a) and \ref{fig:Energy_TDep}b) we present the calculated T-dependence of the interlayer Shear and Breathing mode energies for FLG, along with linear fittings and the corresponding energies obtained from Lattice Dynamics.  Experimental measurements for the 2LG Shear mode by Cong et al.~\cite{Cong2014} are also presented. Even down to T=80 K, the calculated $\omega(T)$ appears linear, which of course contrasts with the real behaviour, as quantum phenomena are absent in MD and more prevalent at low $T$. Also, with increasing $N$, the slopes have distinctly decreasing values for the interlayer modes. For the Shear mode, they go from with $-6.2 \cdot 10^{-3}$ cm$^{-1}$ T$^{-1}$ for bilayer, down to $-4.4 \cdot 10^{-3}$ cm$^{-1}$ T$^{-1}$ for graphite. The same trend is observed for the LBM, where the slopes are ranging from $-9.7 \cdot 10^{-3}$ cm$^{-1}$ T$^{-1}$ for bilayer, down to $-3.6 \cdot 10^{-3}$ cm$^{-1}$ T$^{-1}$ for pentalayer. Again for the LBM of graphite, neither the energies nor the slope follow the above trend, as the energies are higher than those of bilayer, while the slope is between the values of 3ML and 5ML, at $-5.6 \cdot 10^{-3}$ cm$^{-1}$ T$^{-1}$. For both interlayer modes, $\delta\omega_{LD}\approx$ 5-10 \%. To our knowledge, no experimental work on the temperature dependence of the energies of the LBM of FLG has been published so far.

\begin{figure*}[!htb]
    \includegraphics[width=0.85\textwidth]{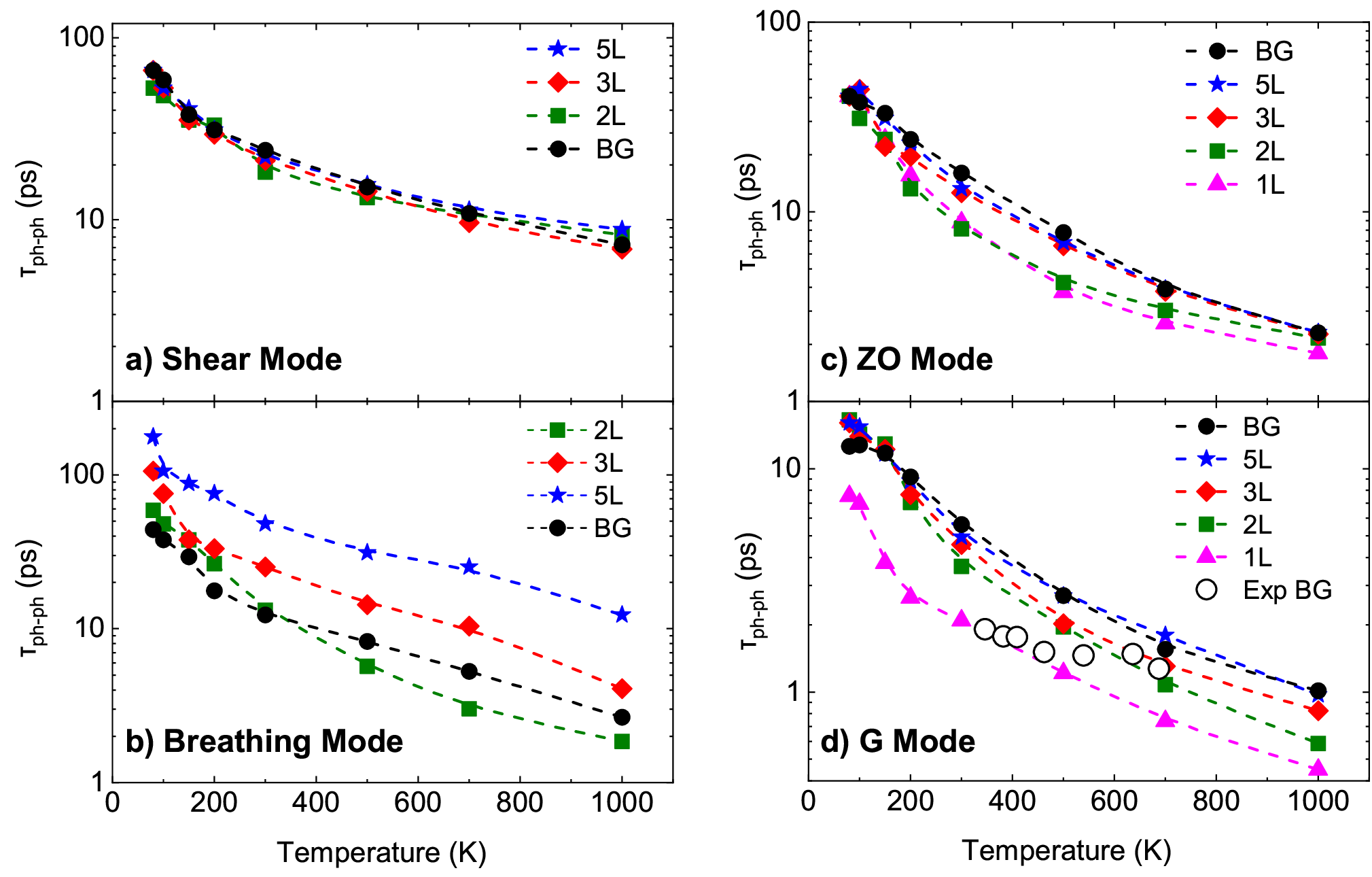}
    \caption{\label{fig:Life_TDep} The calculated temperature dependence of $\tau^{ph-ph}$ of the FLG  (a) Shear, (b) Breathing, (c) ZO and (d) G modes with the number of layers $N$. The correspondence of symbols with $N$ is as follows. 2L: green squares, 3L: red diamonds, 5L: blue stars, BG: black circles. For the G mode, the experimental data from Chatzakis \textit{et al}~\cite{Chatzakis2011} are also presented (hollow circles). Dashed lines connecting points are b-splines as visual guides}
\end{figure*}

A completely different picture is observed for the intralayer modes, where both the energies and the slopes are almost insensitive to $N$. In Figs.~\ref{fig:Energy_TDep}c) and \ref{fig:Energy_TDep}d) we present the temperature dependence of the intralayer ZO and G mode phonon energies for FLG's with different $N$. The energies are slightly upshifted by $\sim$1-5 cm $^{-1}$ going from 1L to graphite. Experimental measurements on the G mode also confirm these findings~\cite{Ferrari2006,Late2011,Tan2012}. The LD energy is not affected by $N$ either. The slopes $\chi_T$ are also practically constant with $N$ and one order of magnitude larger then for the interlayer modes (see Table~\ref{tab:energy_slopes}). Interestingly it is found that for both intralayer modes $\delta\omega_{LD}\approx$ 0.3 \%, which is an order of magnitude lower than $\delta\omega_{LD}$ for the interlayer modes, although on the contrary, the slope $\chi_T$ of the intralayer modes is an order of magnitude higher than $\chi_T$ for the interlayer modes (see Table~\ref{tab:energy_slopes}). 

By comparing with the experimental measurements, we notice again that the Tersoff-2010 potential significantly overestimates the G-mode energies by a factor of  $\sim$5.5 \%, and the slope $\chi_T$ by a factor of  $\sim$2.5~\cite{Yang2020, Liu2019}. Furthermore, the experimental measurements of Yang \textit{et al} in multilayered graphene demonstrated that going from monolayer to FLG's, $\omega_G$ is slightly reduced~\cite{Yang2020}, in contrast to our findings. However, all the measurements in their work were performed on 1ML supported on a Si-substrate. Whether the discrepancy is due to the limitations of the Tersoff potential or to the presence of the substrate in the experimental measurements cannot be determined by this comparison.

Finally the phonon lifetimes due to anharmonicity $\tau^{ph-ph}$ were determined from the linewidth $\Gamma^{ph-ph}$ of the kVACS peaks (FWHM), by the simple relation
\begin{eqnarray}
    \tau^{ph-ph} &=& \frac{\hbar}{\Gamma^{ph-ph}}  \label{eq:Life_G}
\end{eqnarray}
For convenience with units, we use $\hbar=5.29$ ps$\cdot$cm$^{-1}$. In Fig. \ref{fig:Life_TDep}a) and \ref{fig:Life_TDep}b), we present the calculated temperature dependence of lifetimes of the interlayer Shear and Breathing modes respectively, for various numbers of layers. Concerning the Shear mode, it appears that within the uncertainties of the peak fitting, $\tau^{ph-ph}$ is practically insensitive to $N$. This is in agreement with some reported DFT calculations and experimental measurements, where it is suggested that the e-ph contribution in $\tau_C$ is independent of the number of layers, and the measured lifetimes changed very little with $N$~\cite{Tan2012}. The same authors have claimed that the the very low C/G peak Raman intensity ratio and their DFT calculations also indicate that the e-ph contribution to the C-band linewidth is negligible and independent of $N$ therefore the observed Raman linewidth could be directly associated with $\gamma^{ph-ph}_{C}$~\cite{Tan2012}. Other authors have calculated a dominant e-ph contribution and the measured C-band linewidth appears decreasing with temperature~\cite{Cong2014}, which is a sign of e-ph dominated linewidth. On the contrary, for the LBM $\tau^{ph-ph}$ increases with increasing $N$, with the exception of graphite, which appears to have values very close to 2LG. Again, the graphite LBM is an anti-phase mode and its eigenvectors resemble more those of the bilayer LBM, which could also explain the similar $\tau^{ph-ph}(T)$ trends of the two systems. This could also explain the increase of $\tau_{LBM}$ with increasing $N$, since the eigenvectors for the true LBM are mostly zero for the inner layers and mostly the outer layers vibrate. For graphite all layers vibrate, leading to more scattering of the LBM.

Finally, the calculated lifetimes of the intralayer modes are presented in Fig.~\ref{fig:Life_TDep}c) and~\ref{fig:Life_TDep}d) for ZO and G, respectively. Interestingly, for both modes the lifetimes increase with increasing $N$ along the entire temperature range, with 1LG having $\tau_G^{ph-ph}$=1.9~ps while the calculated $\tau_G^{ph-ph}$=5.6~ps, more than double that of the monolayer. Although to our knowledge no experimental study has so far directly measured the $\tau^{ph-ph}_G (T)$ of high-quality, monocrystalline suspended 1L graphene, one would expect intuitively that the weak interlayer interaction in FLG should not significantly affect phonon lifetimes, compared to the monolayer.

Experimentally, the total $\gamma_G$ of suspended 1L graphene at T= 300 K has been measured at 16 cm$^{-1}$~\cite{Lin2011}, giving an estimated value of the total lifetime $\tau_G$= 0.33~ps. For graphite, $\tau^{ph-ph}_{G}(T)$ has been experimentally measured at around $\sim$2.5~ps by multiple authors~\cite{Chatzakis2011,Yan2009}, which is almost 2 times lower than our calculated values with the KC potential ($\sim$5.6~ps). For supported 1L samples, values of $\sim$1$\hyphen$1.5~ps have been reported, increasing with increasing the number of layers $N$~\cite{Kang2010, Katsiaounis2021}, while some authors have measured $\tau^{ph-ph}_{G}\hyphen$2.5~ps, equal to that of graphite~\cite{Wang2010}. Values for 1L suspended on colloidal solutions as high as $4.9 \ ps$ for $\tau^{ph-ph}_{G}$ have also been reported~\cite{Kumar2009}. 

\begin{figure}[!htb]
    \includegraphics[width=0.85
    \columnwidth]{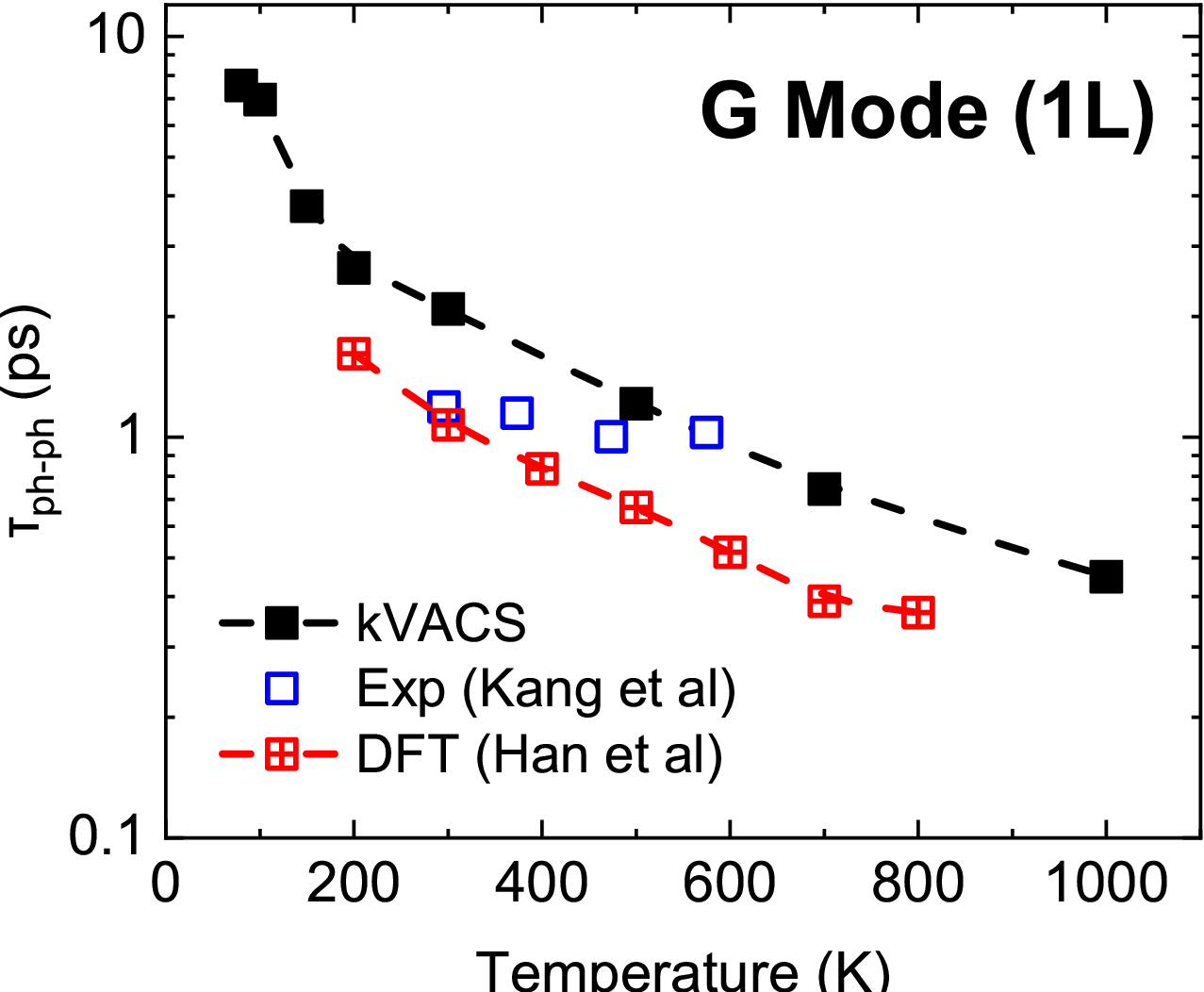}
    \caption{\label{fig:Life_1L} The calculated temperature dependence of $\tau^{ph-ph}$ of the monolayer Raman-active G phonon (black triangles). DFT calculations including 3ph+4ph processes and phonon renormalizaton~\cite{Han2022} are also shown (crossed blue triangles). Calculations are compared with experimental data from TRIARS measurements of supported 1L graphene \cite{Kang2010} (empty red triangles). Dashed lines connecting points are b-splines as visual guides}
\end{figure}

\begin{figure*}[!htb]
    \includegraphics[width=0.85\textwidth]{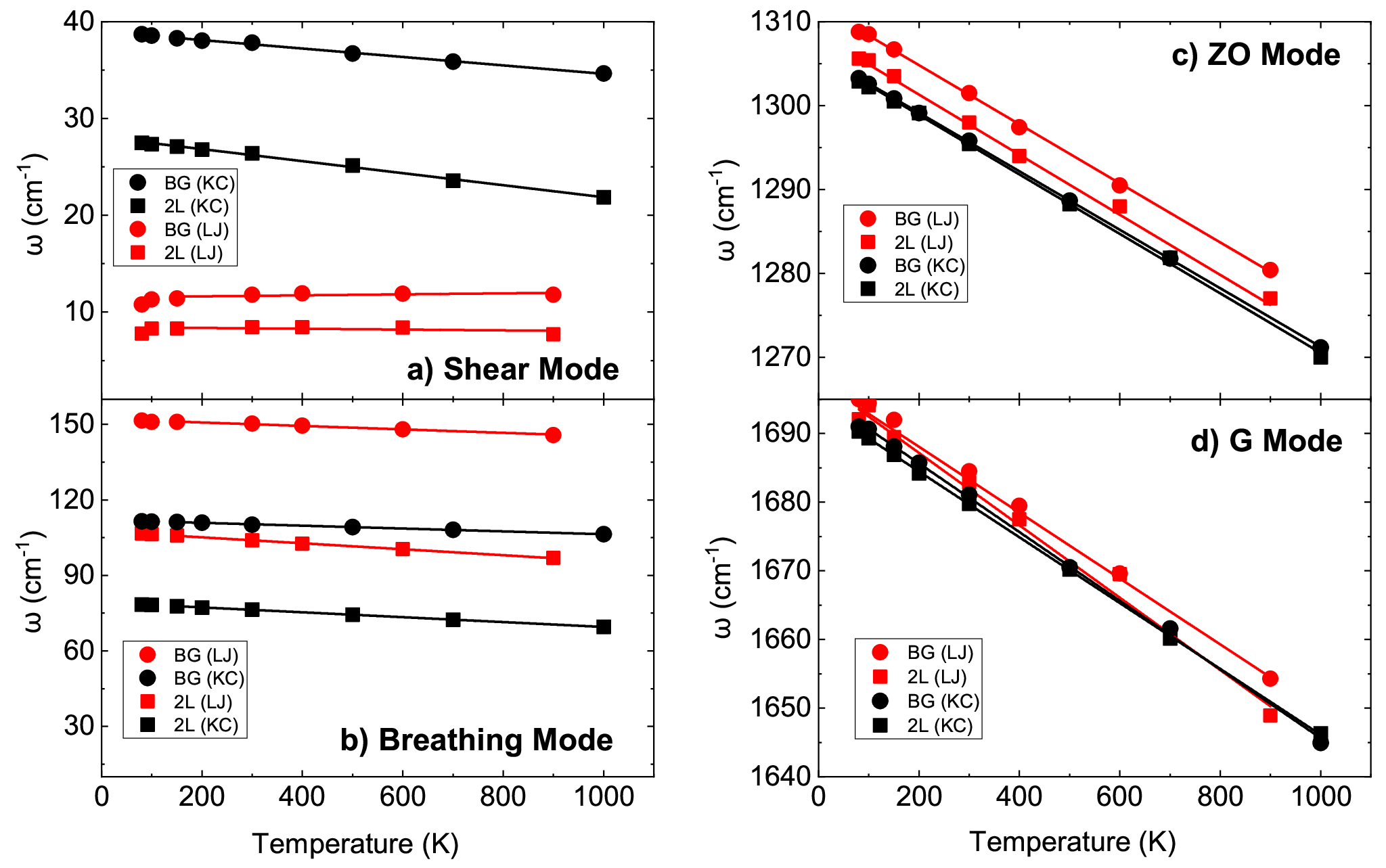}
    \caption{\label{fig:Comp_Energy} The calculated temperature dependence $\omega(T)$ of phonon energies for the interlayer (a) Shear (b) Breathing (c) ZO and (d) G mode in 2L (square symbols) and graphite (round symbols), using KC (black full symbols) and LJ (red crossed symbols) as interlayer potentials.\\}
\end{figure*}

We have also compared our calculations for $\tau^{ph-ph}_G$ with DFT calculations and some of the existing experimental data, shown in Fig.~\ref{fig:Life_1L}. The DFT calculations are for free 1LG and are extracted from the work of Han \textit{et al} (2022) by using only the $\Gamma^{ph-ph}$ from the renormalized 3ph+4ph processes~\cite{Han2022}, and by inputing it in eq.~(\ref{eq:Life_G}). The experimental data are taken from TRIARS measurements of supported 1L graphene by Kang \textit{et al} (2010)~\cite{Kang2010}. The kVACS and DFT curves agree well with each other, although the DFT absolute values are systematically $\sim40 \%$ lower. The experimental values exhibit a different trend, with $\tau^{ph-ph}_G$ remaining almost constant for a temperature $300 K\leq T \leq 600 K$. The authors have also reported that $\tau^{ph-ph}_G$ values for 1L are significantly lower than for graphite, and an increasing trend with increasing $N$, similar to the one observed in Fig. \ref{fig:Life_TDep}b). However, they have attributed this behaviour to the effect of the substrate~\cite{Kang2010}.

In order to clarify the trend of $\tau^{ph-ph}(T)$, we have repeated part of the calculations replacing KC with the Lennard-Jones (LJ) potential for the interlayer interaction. Namely, the T-dependence of the $\Gamma$-point phonon energies and lifetimes for 2L and bulk graphite were repeated, for $80 K\leq T \leq 900 K$. The slopes $\chi_T$ were also calculated and are presented in Table~\ref{tab:energy_slopes_comparison}. To ensure that the trends for both potentials are within the linear regime, the least-square fitting was performed for T $>$ 200 K  for both potentials.

\begin{table}[h]
    \caption{\label{tab:energy_slopes_comparison} The comparison of energy slopes with temperature $\chi_T$ for each phonon mode and for bilayer graphene and graphite, between the two interlayer potentials used (Lennard-Jones and Kolmogorov-Crespi)}
    \begin{ruledtabular}
        \begin{tabular}{ccccc}
        \multirow{2}{*}{($\times$10$^{-3}$ cm$^{-1}$K$^{-1}$)}&\multicolumn{2}{c}{\textbf{KC}}&\multicolumn{2}{c}{\textbf{LJ}}\\
        & Bilayer & Graphite \ & Bilayer \ & Graphite \ \\ \hline
         Shear& -6.2 & -4.3 & -0.4 & -0.5 \\
         LBM& -9.7 & -5.6 & -11.8 & -6.8 \\
         ZO& -35.4 & -34.8 & -35.8 & -35.2 \\
         G& -47.9 & -50.0 & -52.7 & -47.9 \\  
        \end{tabular}
     \end{ruledtabular}
\end{table}

\begin{figure*}[!ht]
    \includegraphics[width=0.85\textwidth]{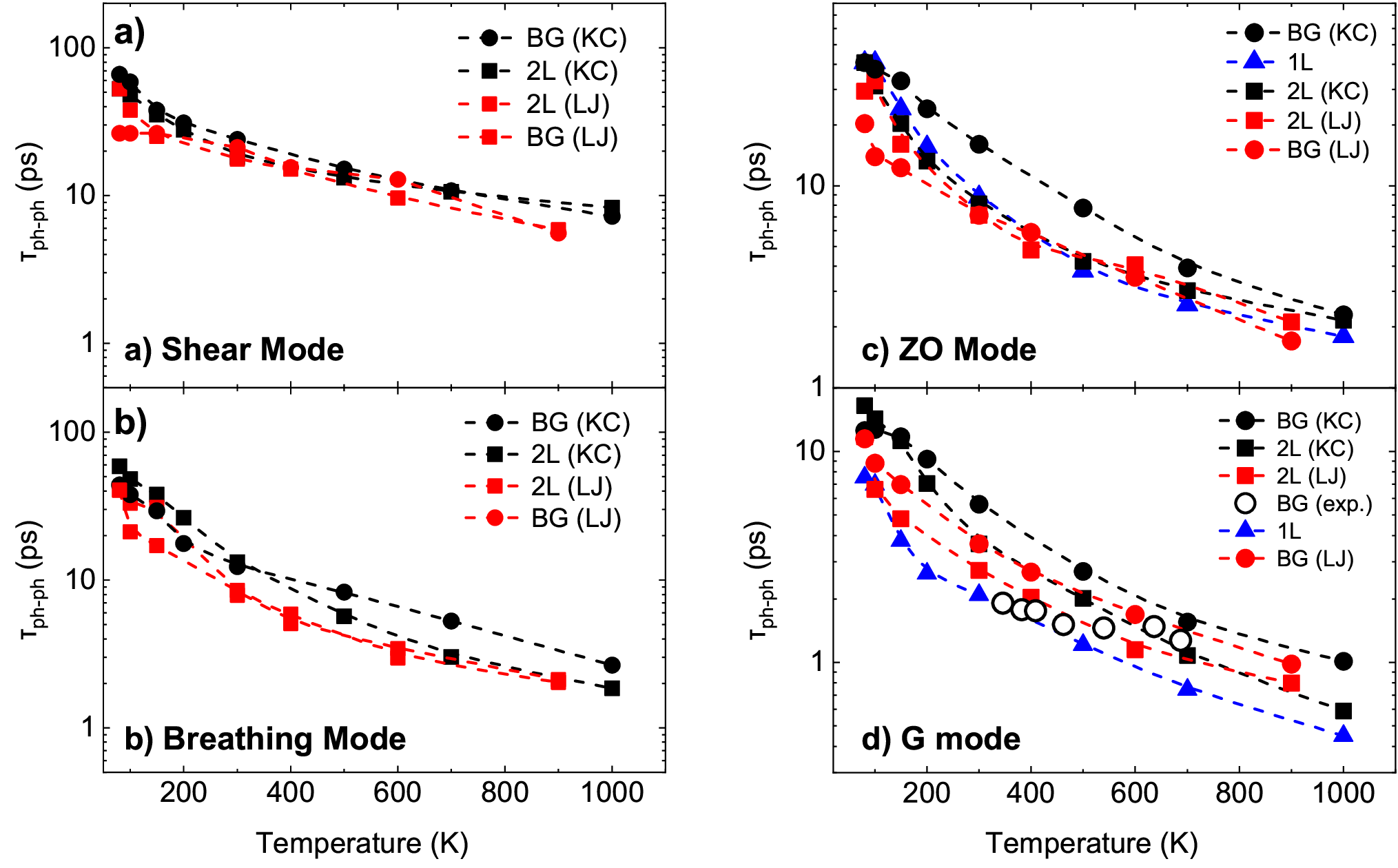}
    \caption{\label{fig:Comp_Life} The calculated temperature dependence $\tau^{ph-ph}(T)$ of phonon lifetimes for the interlayer (a) Shear (b) Breathing (c) ZO and (d) G mode in 2L (square symbols) and graphite (round symbols), using KC (black full symbols) and LJ (red crossed symbols) as interlayer potentials. For the G-mode, the experimental data for graphite from Chatzakis \textit{et al} \cite{Chatzakis2011} are also presented (hollow circles). Dashed lines connecting points are b-splines as visual guides.}
\end{figure*}

In Figs.~\ref{fig:Comp_Energy}a) and \ref{fig:Comp_Energy}b) we present the comparison for the interlayer mode energies, where the trends of the phonon energies $\omega_{inter}(N)$ with increasing number of layers is as expected for both potentials (increasing for the Shear mode and decreasing for LBM). However the absolute values obtained with LJ are quite poor. The LBM energy is significantly overestimated (at $T=300 K$ the error is $30-35\%$, compared to only $0.5-5\%$ for KC), while on the contrary the Shear mode is enormously underestimated (at $T=300 K$ the difference with experiment is $75\%$, compared to only $15\%$ for KC), both for 2L and graphite. Also, the LJ gives for the Shear mode a slope $\chi_T$ that is an order of magnitude lower than the one by KC and, as this slope is already low for KC ($\sim 10^{-3} $ cm$^{-1}$K$^{-1}$), LJ essentially predicts that the Shear mode energy is constant with temperature. The comparison of the slopes for all modes are presented in Table~\ref{tab:energy_slopes_comparison}.

As for the intralayer mode energies, presented in Figs.~\ref{fig:Comp_Energy}c) and \ref{fig:Comp_Energy}d), the differences between the two potentials are much less pronounced, as the intralayer modes are little affected by the weak interlayer interactions. The slopes $\chi_T$ calculated with both potentials agree very well with each other, as can be clearly seen in Table~\ref{tab:energy_slopes_comparison}.  A small but noticeable difference is that for the ZO mode, the difference in energies between 2L and graphite is much more important with LJ ($\sim 3\cdot10^{-3}$) than with KC ($\sim 2\cdot10^{-4}$). This and the overestimation of the LBM mode energies could possibly be related, indicating that the shear and perpendicular interlayer forces cannot be simultaneously fitted in a 2-parameter and 2-body potential like LJ, as the perpendicular components are clearly overestimated, while the shear components are underestimated.

Subsequently, we shall discuss the impact of the choice interlayer potentials on phonon lifetimes with respect to the number of layers $N$. In Figs.~\ref{fig:Comp_Life}a) and \ref{fig:Comp_Life}b) we present this comparison for the interlayer mode lifetimes. It shall first of all be noted that the Shear mode SED peaks exhibit a poor Lorentzian shape and therefore the lifetimes $\tau^{ph-ph}_C(T)$ calculated with LJ have large uncertainties, but are nonetheless presented for reasons of consistency. The LJ results (Fig.~\ref{fig:Comp_Life}a)), however, also demonstrate that for the Shear mode $\tau^{ph-ph}(T)$ is, within the uncertainty of the fitting, practically unaffected by $N$, as was also seen with KC. In terms of absolute values, however, the LJ potential systematically underestimates lifetimes by a factor of 2 compared to KC. As for the LBM lifetimes (Fig.~\ref{fig:Comp_Life}b), again LJ gives systematically underestimates $\tau^{ph-ph}_{LBM}(T)$ and predicts that for graphite and 2L the LBM lifetimes have practically the same value.

Finally, we shall focus on the comparison of the ZO and G mode lifetimes, presented in Figs.~\ref{fig:Comp_Life}c) and \ref{fig:Comp_Life}d) respectively. Beginning with the ZO mode, one may first of all observe that $\tau^{ph-ph}_{ZO}(T)$ for 1L and 2L are practically equal, irrespectful of the interlayer potential used. However, as for the case of the interlayer modes discussed above, LJ gives lower values than KC and predicts that the values for graphite are equal to those of 1L and 2L, in contrast to KC, which predicts that $\tau^{BG}_{ZO}>\tau^{1L}_{ZO}, \tau^{2L}_{ZO}$

Last, focusing on the G mode lifetimes, one can make the following observations: Both potentials give results that are close in terms of absolute value and with respect to the number of layers $N$, in general agreement with experiment. Namely, both potentials predict a clear monotonous increase of $\tau^{ph-ph}_{G}$ with $N$, although again LJ predicts lower absolute values.  Also, for temperatures close to room temperature, the LJ results for 1L and graphite agree closer with TRIARS experimental data by Chatzakis \textit{et al} for graphite~\cite{Chatzakis2011}, all giving a value of approximately $\tau^{ph-ph}_{G}(300 K)\approx 2$ ps, while the KC predicted value for graphite is more than two times larger (5.6 ps). In any case, these results confirm that with increasing the number of layers, the G-mode lifetime also increases, although the amount of increase depends on the type of interlayer potential used.

\begin{figure}[!htb]
    \includegraphics[width=0.85\columnwidth]{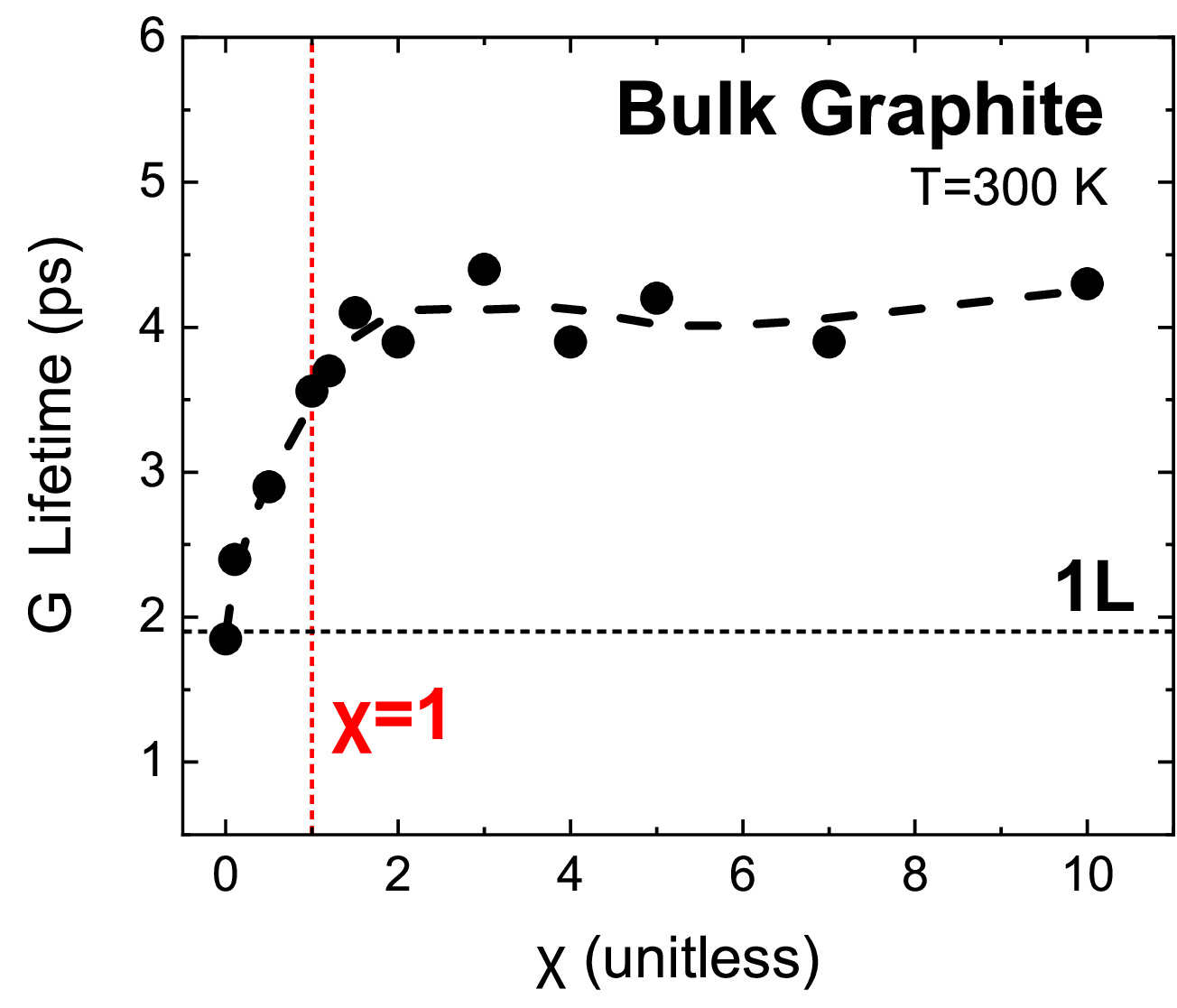}
    \caption{\label{fig:Couple_Strength} The effect of the coupling strength $\chi$ on the G-mode lifetimes for BG calculated with the LJ potential. The horizontal line (black) represents the calculated value for the 1LG. The vertical line (red) represents the value $\chi$ = 1, which corresponds to the normal coupling strength that gives $\tau_G$ = 3.6 ps. Dashed lines connecting points are b-splines as visual guides. }
\end{figure}

To conclude this work, and in order to shed further light on how the nature of the interlayer interaction affects the lifetimes in FLG, we have performed further calculations on the G-mode lifetime of Bulk Graphite, using the LJ potential modified with a coupling strength factor $\chi$, which simply multiplies the LJ energy by $\chi$. We have calculated the lifetime for BG with $\chi$ ranging from 0 (non-interacting single layers) to 10 (strongly interacting layers) and present it in Fig.~\ref{fig:Couple_Strength}. There are two main points that can be made from these results. First of all, increasing $\chi$ increases $\tau_{G}$, and for large enough $\chi>2$, the lifetime reaches a plateau of $\sim$4~ps, which could be understood as the strong-coupling limit. Secondly, the increae of the BG lifetimes with the coupling strength may indicate that the increasing trend of $\tau_G$ with $N$ is due to an increase of the interlayer interaction strength with $N$, as with increasing $N$ each layer interacts with more non-adjacent layers (mainly through the long-range part of the interlayer VdW interaction).

\section{\label{sec:Conclusions}Conclusions:}

This study presents a comprehensive systematic analysis of the phonon properties of Few-Layered Graphene. We explore the effects of the number of graphene layers, which has not been comprehensively studied computationally before, as well as the temperature dependence of phonon properties, including the first-time calculation of the temperature dependence of the Layer Breathing Mode (LBM) energy. Furthermore, we investigate the influence of the type of interlayer potential and the strength of interlayer coupling on phonon properties. Our investigation focuses particularly on phonon lifetimes, which present both experimental and computational challenges.

The behavior of inter- and intra-layer phonon modes exhibits notable differences. Inter-layer phonon energies are significantly more affected by the presence and type of interlayer interactions. As for the lifetimes, both intra-layer and LBM lifetimes increase with the number of layers, whereas for the Shear mode they are unaffected by the number of layers. One notable finding is that the Raman-active G-mode phonon lifetime $\tau_G$ increases with the number of layers $N$, regardless of the type of interlayer potential. This dependence is more pronounced with the use of the Kolmogorov–Crespi (KC) potential, while the Lennard-Jones (LJ) potential yields systematically lower values, potentially indicating increased scattering.  The effect of interlayer coupling strength on G-mode lifetimes was also investigated. Increasing the LJ interlayer coupling strength leads to an increase in $\tau_G$ for graphite, up to $\chi$=2, where it reaches a plateau that indicates a strong-coupling threshold. This observation suggests that the increase in $\tau_G$ with the number of layers may be attributed to the increasing interlayer coupling strength with the addition of more layers.

The Kolmogorov–Crespi (KC) potential was proven to be highly effective in describing interlayer interactions for multi-layered graphene, particularly for the shear forces, which are challenging to model with simpler potentials. It manages to fairly reproduce the experimental Shear and Breathing mode energies, although it tends to overestimate the lifetimes of the Raman-active G-mode in graphite by a factor of 2. Conversely, the Lennard-Jones (LJ) potential performs poorly in describing C-C interlayer interactions. It fails to simultaneously capture both shear and perpendicular interlayer forces. Particularly for the Shear mode, the LJ potential exhibits undesirable behavior, including the prediction of extremely low values of energy, which practically remained constant with temperature, and a poor Lorentzian shape of the Spectral Energy Density, leading to significant uncertainties in phonon lifetimes.

We advocate for further research combining Density Functional Theory (DFT) simulations with experimental studies to explore the effect of the number of graphene layers on phonon lifetimes due to phonon-phonon interactions. This includes DFT simulations in multilayer graphene and Time-Resolved Incoherent Anti-Stokes Raman Spectroscopy (TRIARS) experiments in high-quality suspended and monocrystalline FLG, in order to isolate the anharmonic contribution to the phonon lifetime.

\section*{\label{sec:Vide} Acknowledgments}
The major part of the calculations presented in this work was performed in the \textit{Jean-Zay} CPU array of the IDRIS High Performance Computing facilities (HPC) under the allocation 2022-AD010913913 made by
GENCI. Part of the calculations was also performed in the \textit{Aristotelis} HPC facilities of the Aristotle University of Thessaloniki. ENK acknowledges support for part of this ressearch project in the framework of H.F.R.I call “Basic research
Financing (Horizontal support of all Sciences)” under the National Recovery and Resilience
Plan “Greece 2.0” funded by the European Union - NextGenerationEU (H.F.R.I. Project Number: 15932).

\section*{\label{sec:Appendices}Appendices:}
\appendix
\section{The Kolmogorov-Crespi (KC) interlayer potential:} \label{app:KC}
The KC potential is given by~\cite{Kolmogorov2005}
\begin{eqnarray*}    
    V(\textbf{r}_{ij}, \textbf{n}_i, \textbf{n}_j) &=& e^{-\lambda(r_{ij}-z_0)}\left[C + f(\rho_{ij}) + f(\rho_{ji})\right] \\
    &-& A\left(\frac{r_{ij}}{z_0}\right)^{-6},
\end{eqnarray*}
\begin{eqnarray*}
    \rho_{ji}&=&r_{ij} - \left( \textbf{n}_i \textbf{r}_{ij}\right)^2
\end{eqnarray*}
\begin{eqnarray}
    f(\rho) &=& e^{-(\rho / \delta)^2} \sum C_{2n} (\rho/\delta)^{2n} \label{eq:KC}.
\end{eqnarray}

\noindent where $\textbf{n}_i$ is the local normal to the $sp^2$ plane in the vicinity of atom $i$, pointing to the direction of the $p_z$ orbitals of atom $i$ and $\rho_{ji}$ the transverse distance of the $p_z$ orbitals of atoms $i$ and $j$ belonging to different layers. $C$, $\lambda$, $z_0$, $A$, $\delta$ and $C_{2n}$ are fitting parameters. Upon reparametrization, the KC potential has been shown to be able to reproduce very accurately the interlayer binding energy curve and sliding energy surfaces of bilayer graphene \cite{Ouyang2018}. 

\section{The k-Space Velocity Autocorrelation (kVACS) method:} \label{app:kVACS_Method}
The central quantity that is calculated with this method is the Velocity Autocorrelation Sequence (k-VACS) denoted by $Z_x^p(\textbf{k},t_i)$. In order to arrive at its definition, we first define the k-space velocity Fourier transform $v_x^p(\textbf{k},t)$ given by 
\begin{eqnarray}
    v_x^p(\textbf{k},t) &=& \frac{1}{\sqrt{N}} \sum_{l=1}^{N} v_x(\textbf{r}_l^p,t) e^{-i\textbf{k}\cdot\textbf{R}_l} \label{eq:vk}
\end{eqnarray}
where $x$ denotes the cartesian component, $\textbf{r}_l^p$ is a label for the atom species labeled $p$ inside the primitive unit cell labeled $l$, and $v_x(\textbf{r}_l^p,t)$ its respective velocity at time $t$. Also $\textbf{R}_l$ marks the position of the primitive unit cell $l$ within the computational cell. The kVACS $Z_x^p(\textbf{k},\omega)$ is then defined as a correlation function
\begin{eqnarray}
    Z_x^p(\textbf{k},\omega) &=& \int d\tau e^{i\omega \tau}\langle v_x^{*p}(\textbf{k},\tau)v_x^{p}(\textbf{k},0) \rangle  \label{eq:Z_t}.
\end{eqnarray}
where the brackets $\langle \cdots \rangle$ denote time averaging, the asterisk $^*$ denotes complex conjugation and $\tau$ is the correlation time. It should be noted that normally the angular brackets denote ensemble averages, which are not directly computable in MD simulations, therefore we use time-averages instead. Since at the time scales of the samples we obtained (in the order of 350 ps) ergodicity is not guaranteed, we perform multiple independent simulations which we average to obtain the final $Z_x^p(\textbf{k},\omega)$. By adding the contributions of all atom species $p$ we obtain the polarised kVACS $Z_x(\textbf{k},\omega)$, and again by summing over all polarizations $x$, we get the total 
\begin{eqnarray}
    Z_{total}(\textbf{k},\omega) &=& \sum_x \sum_p m_pZ_x^p(\textbf{k},\omega) \label{eq:Z_total}
\end{eqnarray}
In Appendix \ref{app:kVACS_proof} we show that the kVACS is approximately a superposition of Lorentzian peaks having a central positions and FWHM's equal to the energies $\omega_{\textbf{k},j}$ and scattering rates $2\Gamma_{\textbf{k},j}$ respectively of the phonon mode $j$ at point $\textbf{k}$, given by
\begin{eqnarray}
    Z_{total}(\textbf{k},\omega) &=& \sum_{j} \frac{A_j}
    {\left(\omega-\omega_{\textbf{k},j}\right)^2 + \Gamma_{\textbf{k},j}^2}  \label{eq:Lor_Superposition}
\end{eqnarray}

\noindent which is exactly the phonon Spectral Energy Density (SED) of the system. As it has been already mentioned before \cite{Koukaras2015}, taking the velocity as a Wide-Sense Stationary (WSS) random process, we can apply the Wiener-Khintchine theorem \cite{Blackman1959} to prove that $Z_x^p(\textbf{k},\omega) \propto \Phi_x^p(\textbf{k},\omega)$, where $\Phi_x^p(\textbf{k},\omega)$ is the phonon SED, given by \cite{Thomas2010}
\begin{eqnarray}
    \Phi_x^p(\textbf{k},\omega) &=& \left| \int dt e^{i\omega t} v_x^p(\textbf{k},t)\right|^2 \nonumber \\
    &\coloneqq& \left|v_x^p(\textbf{k},\omega)\right|^2 \label{eq:SED}
\end{eqnarray}
Again, by summing over all polarizations $x$ and atom species $p$, we get the total $\Phi(\textbf{k},\omega) = \sum_{x, p} m_p \Phi_x^p(\textbf{k},\omega)$. The conditions for a random process to be a WSS is to be square integrable, having a stationary mean value and an autocorrelation function that does not depend on the reference time $t_0$, which are in general fulfilled for the velocity of a stationary system in thermal equilibrium. What is more, calculating $\Phi_x(\textbf{k},\omega)$ instead of $Z_x(\textbf{k},\omega)$ is computationally more efficient, as the later via eq.~(\ref{eq:Z_t}) is of order $\mathcal{O}(N^2)$, while the the SED in eq.~(\ref{eq:SED}) is of order $\mathcal{O}(NlogN)$, with respect to the total number $N$ of sampled time points.

Finally, $\Phi_x(\textbf{k},\omega)$ can be related to the phonon Density of States (pDOS) $g(\omega)$ by the relation \cite{Chakraborty2018}
\begin{eqnarray}
    &&g(\omega) = \sum_x \sum_p m_p \int \Phi_x^p(\textbf{k},\omega) d\textbf{k} \nonumber \\
    &&= \frac{1}{3k_B T}\sum_x \sum_p m_p \sum_l \left|\int dt e^{i\omega t}v_x(\textbf{r}_l^p,t) \right|^2  \label{eq:pDOS}
\end{eqnarray}

\section{Proof that $Z_i(\textbf{k}, \omega) \propto \frac{1}{\left( \omega-\omega_{i,0} \right)^2 + \Gamma_i^2}$} \label{app:kVACS_proof}
In this appendix we shall prove that the kVACS $Z_i(\textbf{k}, \omega)$ is approximately a superposition of Lorentzian peaks having a central positions and FWHM's equal to the energies $\omega_{\textbf{k},j}$ and scattering rates $2\Gamma_{\textbf{k},j}$ respectively of the phonon mode $(\textbf{k},j)$ where $\textbf{k}$ is the wavevector and $j$ is the phonon branch. First, we shall recall from basic quantum mechanics that the velocity operator $\textbf{v}_l$ in the Heisenberg representation is given by 
\begin{eqnarray}
    \textbf{u}_l(t) &\coloneq& e^{i\frac{H}{\hbar}t} \textbf{u}_l e^{-i\frac{H}{\hbar}t} \\
    \textbf{v}_l(t) &\coloneq& \frac{\textbf{p}_l(t)}{m} = \frac{\partial \textbf{u}_l(t)}{\partial t} = -\frac{i}{\hbar}\left[\textbf{u}_l(t), H\right] \label{eq:vel_heisenberg}
\end{eqnarray}
\noindent where $\textbf{p}_l(t)$ is the momentum operator, $\textbf{u}_l(t)$ is the displacement operator and $H$ is the Hamiltonian of the system. The subscript index $l$ is an index of atoms. As a corollary, the spatial Fourier Transform (FT) of the velocity operator is given by
\begin{equation}
    \textbf{v}(\textbf{k},t) \coloneq \sum_l \textbf{v}_l(t)e^{-i\textbf{k} \cdot \textbf{R}_l} = -\frac{i}{\hbar}\left[\textbf{u}_l(\textbf{k},t), H\right]
    \label{eq:vel_fourier}
\end{equation}
\noindent where $\textbf{R}_l$ is the equilibrium position of atom $l$, and is not an operator. Also $\textbf{u}_l(\textbf{k},t)$ is the spatial FT of the displacement operator $\textbf{u}_l(t)$. Now, we shall focus on the treatment of the main quantity $Z_{xy}(\textbf{k},\omega)$. As per eq.~(\ref{eq:Z_t}),  it is defined as
 \begin{eqnarray}
    Z_{xy}(\textbf{k},\omega) &=& \int d\tau e^{i\omega \tau} \langle v_x^{\dagger}(\textbf{k},\tau)v_y(\textbf{k},0) \rangle \nonumber \\
    &=&\int d\tau e^{i\omega \tau} \langle v_x(-\textbf{k},\tau)v_y^{p}(\textbf{k},0) \rangle
    \label{eq:Z_omega}
\end{eqnarray}
\noindent where $\tau$ is the correlation time. The second line results from the relation $A^{\dagger}(\textbf{k},t) = A(-\textbf{k},t)$ which is true for a Hermitian operator $A_l(t)$ depending on the atom index $l$. We shall finally use the quantity $B_{xy}(\textbf{k}, \omega)$ defined as
\begin{equation}
    B_{xy}(\textbf{k}, \omega)= \int d\tau e^{i\omega \tau} \langle u_x(-\textbf{k},\tau)u_y(\textbf{k},0) \rangle
    \label{eq:B_xy}
\end{equation}
\noindent This quantity is closely related to the phonon propagator, which gives the phonon self-energy $\Sigma_{\textbf{k}j}(\omega)=\Delta_{\textbf{k}j}(\omega) + i \Gamma_{\textbf{k}j}(\omega)$ \cite{Mahan1990}. In the seminal work by Maradudin and Fein (1962), it was proven that if the anharmonic terms in the crystal potential are small enough to be considered as perturbations, that is $\Delta_{\textbf{k}j}, \Gamma_{\textbf{k}j} \ll \omega_{\textbf{k}j}$, then they can both be considered as independent of $\omega$, giving~\cite{Maradudin1962}
\begin{eqnarray}
    B_{xy}(\textbf{k}, \omega) &=& \frac{1}{Nm} \left(\frac{\hbar}{1-e^{-\beta \hbar \omega}}\right) \sum_j \left[\frac{e^*_x(\textbf{k},j)e_y(\textbf{k},j)}{\omega(\textbf{k},j)}\right] \nonumber \\
   &\times& \frac{\Gamma_{\textbf{k}j}}{\left(\omega-\omega(\textbf{k},j)- \Delta_{\textbf{k}j}\right)^2+ \Gamma^2_{\textbf{k}j}}
   \label{eq:B_xy_Lorentzian}
\end{eqnarray}
\noindent where $N$ is the number of unit cells, $m$ is the atomic mass, $e_x(\textbf{k},j)$ is the $x$-coordinate of the eigenvector of the phonon mode $j$ at point $\textbf{k}$, $\beta=1/k_BT$ and $k_B$ is the Boltzmann constant.

We shall now proceed to prove that $Z_{xy}(\textbf{k},\omega)=\omega^2B_{xy}(\textbf{k}, \omega)$. First, we recall that the thermodynamical average in the canonical ensemble of an operator $X$ is given by \cite{Mahan1990}
\begin{equation}
    \langle X \rangle = \frac{1}{Z} \sum_a e^{\beta E_a}\langle a|X|a \rangle
    \label{eq:thermo_average}
\end{equation}
\noindent where $Z$ is the partition function, $|a\rangle$ is the eigenstate of the Hamiltonian with energy $E_a$ and $k_B$ is the Boltzmann constant. Therefore, starting from eq.~(\ref{eq:Z_omega}) and applying the resolution of identity, we have
\begin{eqnarray}
    &Z_{xy}&(\textbf{k},\omega) = \frac{1}{Z} \sum_{a}e^{\beta E_a} \sum_{\beta} \nonumber \\
    &\times& \int d\tau e^{i\omega \tau} \langle a| v_x(-\textbf{k},\tau)|\beta \rangle \langle \beta |v_y(\textbf{k},0)|a \rangle
    \label{eq:Z_omega_2}
\end{eqnarray}
\noindent Now, focusing only on the integral of the second line in eq.~(\ref{eq:Z_omega_2}), if we expand the commutator of eq.~(\ref{eq:vel_fourier}) and apply the Hamiltonian on the eigenstates, we get
\begin{flalign}
    \left(-\frac{i}{\hbar}\right)^2 \int d\tau e^{i\omega \tau} & \langle a| u_x(-\textbf{k},\tau)H-Hu_x(-\textbf{k},\tau)|\beta \rangle &&\nonumber \\
    &\times \langle \beta |u_y(\textbf{k},0)H-Hu_y(\textbf{k},0) |a\rangle &&\nonumber
\end{flalign}
\begin{flalign}
    =\left(-\frac{i}{\hbar}\right)^2 \int d\tau e^{i\omega \tau} &\langle a | u_x(-\textbf{k},\tau)|\beta \rangle \langle \beta |u_y(\textbf{k},0) |a \rangle &&\nonumber \\
    \times & \left[-\left(E_a-E_{\beta}\right)^2\right] && \nonumber
\end{flalign}
\begin{flalign}
    =-\left[\frac{i(E_a-E_{\beta})}{\hbar}\right]^2 \int d\tau e^{i\omega \tau} &\langle a| e^{i\frac{H}{\hbar}\tau} \textbf{u}_x(-\textbf{k}) e^{-i\frac{H}{\hbar}\tau}|\beta \rangle && \nonumber\\ 
    \times &\langle \beta |u_y(\textbf{k},0) |a \rangle && \nonumber
\end{flalign}
\begin{flalign}
    =\left(\frac{E_{\beta}-E_a}{\hbar}\right)^2 & \langle a|\textbf{u}_x(-\textbf{k})|\beta \rangle \langle \beta |u_y(\textbf{k}) |a\rangle &&\nonumber \\
    \times &\int d\tau e^{i\left(\omega - \frac{E_{\beta}-E_a}{\hbar}\right)}  &&
    \label{eq:Z_integral_delta}
\end{flalign}
\begin{flalign}
    =\omega^2 \langle a| \textbf{u}_x(-\textbf{k})|\beta \rangle \langle \beta |u_y(\textbf{k}) &|a\rangle  && \nonumber \\ 
    \times &\delta\left(\omega - \frac{E_{\beta}-E_a}{\hbar}\right) &&
    \label{eq:Z_integral}
\end{flalign}

\noindent As the integral in eq.~(\ref{eq:Z_integral_delta}) is a Dirac delta function, the only non-zero terms in the integral of eq.~(\ref{eq:Z_omega_2}) that survive are those where $\omega= \frac{E_{\beta}-E_a}{\hbar}$, therefore we can directly replace $\left(\frac{E_{\beta}-E_a}{\hbar}\right)^2$ by $\omega^2$, obtaining eq.~(\ref{eq:Z_integral}). From this point it is straightforward to rewrite eq.~(\ref{eq:Z_integral}) backwards as
\begin{equation}
   \omega^2 \int d\tau e^{i\omega \tau} \langle a| u_x(-\textbf{k},\tau)|\beta \rangle \langle \beta |u_y(\textbf{k},0)|a \rangle
\end{equation}
\noindent which is exactly the integral of eq.~(\ref{eq:Z_omega_2}). Finally, as the $\omega^2$ factor stays outside the double summation $\sum_a \sum_{\beta}$, we have that
\begin{flalign}
    Z_{xy}(\textbf{k},\omega) &= \omega^2 \int d\tau e^{i\omega \tau} \langle u_x(-\textbf{k},\tau)u_y(\textbf{k},0) \rangle && \nonumber \\
   & = \omega^2 B_{xy}(\textbf{k}, \omega)
   \label{eq:Z_proof}
\end{flalign}
Therefore, we have proven that relation for the quantum-mechanical correlation functions (CF). The usual `recipe' to obtain the classical CF from the quantum mechanical CF is to apply the limit $\hbar \rightarrow 0$ \cite{Zwanzig1965}. By applying this limit to the expression for $B_{xy}(\textbf{k}, \omega)$ in eq.~(\ref{eq:B_xy_Lorentzian}), we obtain for the prefactor
\begin{equation}
    \lim_{\hbar \rightarrow 0} \frac{\hbar}{1-e^{-\beta \hbar \omega}} = \lim_{\hbar \rightarrow 0} \frac{\hbar}{1-\left(1-\beta \hbar \omega\right)} = \frac{1}{\beta \omega}
    \label{eq:prefactor}
\end{equation}

Another important point to be mentioned is the behaviour of the self energy components $\Delta_{\textbf{k}j}(\omega)$ and $\Gamma_{\textbf{k}j}(\omega)$ in the limit $\hbar \rightarrow 0$. As was pointed out in eqs (5.11 a,b) in Maradudin and Fein (1962) \cite{Maradudin1962}, this limit is equivalent to taking the limit $T \rightarrow \infty$ and is found to be finite and non-zero. Therefore, combining eqs~(\ref{eq:B_xy_Lorentzian}), (\ref{eq:Z_proof}) and (\ref{eq:prefactor}), we can finally write
\begin{flalign}
    Z_{xy}(\textbf{k},\omega) &=  \frac{\omega k_BT }{Nm} \sum_j \left[\frac{e^*_x(\textbf{k},j)e_y(\textbf{k},j)}{\omega(\textbf{k},j)}\right] \nonumber \\
   &\times \frac{\Gamma_{\textbf{k}j}}{\left(\omega-\omega(\textbf{k},j)- \Delta_{\textbf{k}j}\right)^2+ \Gamma^2_{\textbf{k}j}}
   \label{eq:Z_xy_w_Lorentzian}
\end{flalign}
\noindent which is not exactly Lorentzian, as it has a prefactor that depends on $\omega$. However, as per the assumption that $\Delta_{\textbf{k}j}, \Gamma_{\textbf{k}j} \ll \omega_{\textbf{k}j}$, the prefactor is varying much more slowly than the Lorentzians, which almost behave as delta functions, therefore we can approximate it as a constant $\omega \approx \omega_{\textbf{k},j}$, finally getting
\begin{equation}
    Z_{xy}(\textbf{k},\omega) =  \frac{k_BT}{NM} \sum_j 
   \frac{\Gamma_{\textbf{k}j} e^*_x(\textbf{k},j)e_y(\textbf{k},j)}{\left(\omega-\omega(\textbf{k},j)- \Delta_{\textbf{k}j}\right)^2+ \Gamma^2_{\textbf{k}j}}
   \label{eq:Z_xy_Lorentzian}
\end{equation}
\noindent which is now a superposition of Lorentzians, with positions $\left(\omega(\textbf{k},j)+ \Delta_{\textbf{k}j}\right)$ and FWHM $2\Gamma_{\textbf{k}j}$ as was to be proven.

As a sidenote, the above derivation was done for a crystal having a single atom species in its unit cell. It is straightforward to extend this to multiple species, by simply replacing $m \rightarrow m_p$ and $e_x(\textbf{k},j) \rightarrow e_x^p(\textbf{k},j)$, where now $p$ is an index to atom species. Then, by remembering that for each phonon mode $(\textbf{k}, j)$ the eigenvectors are orthonormal $\sum_{x,p} e_x^p(\textbf{k},j)e_x^{*p}(\textbf{k},j) =1$, we can write $Z_{total}(\textbf{k},j)$ as 

\begin{align}
    &Z_{total}(\textbf{k},\omega) = \sum_{x,p} m_pZ_{xx}(\textbf{k},\omega)\nonumber \\
    &=\frac{k_BT}{N} \sum_j \frac{\Gamma_{\textbf{k}j}}{\left(\omega-\omega(\textbf{k},j)- \Delta_{\textbf{k}j}\right)^2+ \Gamma^2_{\textbf{k}j}}
   \label{eq:Z_total_Lorentzian}
\end{align}
where we have used the correspondence between eqs (\ref{eq:Z_total}) and (\ref{eq:Z_omega}) as $Z_{x}(\textbf{k},\omega)=Z_{xx}(\textbf{k},\omega)$

\end{document}